\documentclass[%
reprint,
 amsmath,amssymb,
 aps,
superscriptaddress
]{revtex4-1}

\usepackage{graphicx}
\usepackage{dcolumn}
\usepackage{bm}
\usepackage{bbding}
\usepackage{braket}
\usepackage{physics}
\usepackage{pifont}
\usepackage[colorlinks,
            linkcolor=blue,       
            anchorcolor=blue,  
            citecolor=blue,
            urlcolor=blue,
            ]{hyperref}

\begin{document}

\title{Truncated Atomic Plane Wave Method for the Subband Structure Calculations of Moir\'e Systems}


\author{Wangqian Miao}
\affiliation{Materials Department, University of California, Santa Barbara, California 93106-5050, USA}
\author{Chu Li}
\affiliation{Department of Physics, The Hong Kong University of 
Science and Technology, Clear Water Bay, Hong Kong, China}
\affiliation{HKUST Shenzhen-Hong Kong Collaborative
Innovation Research Institute, Shenzhen, China}
\author{Xu Han}
\affiliation{Department of Physics, The Hong Kong University of 
Science and Technology, Clear Water Bay, Hong Kong, China}
\affiliation{HKUST Shenzhen-Hong Kong Collaborative
Innovation Research Institute, Shenzhen, China}
\author{Ding Pan}
\affiliation{Department of Physics, The Hong Kong University of 
Science and Technology, Clear Water Bay, Hong Kong, China}
\affiliation{Department of Chemistry, The Hong Kong University of Science and Technology, Clear Water Bay, Hong Kong, China}
\affiliation{HKUST Shenzhen-Hong Kong Collaborative
Innovation Research Institute, Shenzhen, China}
\author{Xi Dai}
\email{daix@ust.hk}
\affiliation{Department of Physics, The Hong Kong University of 
Science and Technology, Clear Water Bay, Hong Kong, China}
\affiliation{Materials Department, University of California, Santa Barbara, California 93106-5050, USA}


\date{\today}

\begin{abstract}
We propose a highly efficient and accurate numerical scheme named Truncated Atomic Plane Wave (TAPW) method to determine the subband structure of Twisted Bilayer Graphene (TBG) inspired by the Bistritzer-MacDonald (BM) model. Our method utilizes real space information of carbon atoms in the moir\'e unit cell and projects the full tight binding Hamiltonian into a much smaller subspace using atomic plane waves. Using our new method, we are able to present accurate electronic band structures of TBG in a wide range of twist angles together with detailed moir\'e potential and screened Coulomb interaction at the first magic angle. Furthermore, we generalize our formalism to solve the problem of low frequency moir\'e phonons in TBG.
\end{abstract}

\maketitle

\newcommand{\bk}{\mathbf{k}}
\newcommand{\br}{\mathbf{r}}
\newcommand{\bq}{\mathbf{q}}
\newcommand{\bp}{\mathbf{p}}
\newcommand{\bg}{\mathbf{g}}
\newcommand{\bG}{\mathbf{G}}
\newcommand{\bR}{\mathbf{R}}
\newcommand{\bK}{\mathbf{K}}
\newcommand{\ri}{\mathrm{i}}
\newcommand{\rd}{\mathrm{d}}
\newcommand{\e}{\mathrm{e}}
\newcommand{\bA}{\mathbf{A}}
\newcommand{\I }{\mathrm{I}}
\newcommand{\J }{\mathrm{J}}
\newcommand{\rI}{\mathrm{I}}
\newcommand{\rJ}{\mathrm{J}}
\newcommand{\bL}{\mathbf{L}}
\newcommand{\btau}{\bm{\tau}}
\newcommand{\bH }{\mathbf{H}}
\newcommand{\bkbar}{\bar{\mathbf{k}}}
\newcommand{\bqbar}{\bar{\mathbf{q}}}
\newcommand{\bfk}{\mathbf{k}}
\newcommand{\bfr}{\mathbf{r}}
\newcommand{\bfq}{\mathbf{q}}
\newcommand{\bfp}{\mathbf{p}}
\newcommand{\bQ}{\mathbf{Q}}
\newcommand{\bfu}{\mathbf{u}}
\newcommand{\re}{\mathrm{e}}

\newcommand{\au}{\mathbf{a}^{(2)}}
\newcommand{\bu}{\mathbf{b}^{(2)}}
\newcommand{\ab}{\mathbf{a}^{(1)}}
\newcommand{\bb}{\mathbf{b}^{(1)}}
\newcommand{\bRu}{\mathbf{R}^{(2)}}
\newcommand{\bRb}{\mathbf{R}^{(1)}}
\section{Introduction}

Twisted Bilayer Graphene (TBG), formed by stacking one single layer on the other with a small twist, provides a great platform for physicists 
to study novel quantum phenomena. More attention has been attracted after the discovery of unconventional superconductivity, orbital magnetism and correlated insulating phases
\cite{Cao2018,Cao2018a,Lu2019,Yankowitz2019,Xie2019,Choi2019,Jiang2019,Kerelsky2019,Sharpe2019,Cao2020,Serlin2020,wong_cascade_2020,nuckolls_chern_2020,choi2021correlation,saito2020,das2020symmetry,wu_chern_2020,oh_unconventional_2021}
in TBG systems at the magic angle, around 1.1$^\circ$. The mechanism behind these observations is still an open question.

Unlike aligned bilayer graphene, in TBG systems, a moir\'e pattern forms in the real space which breaks original periodicity of graphene. This pattern leads to the difficulty for band structure calculations due to the huge amount of atoms in a single unit cell. In order to overcome this kind of difficulty, several low energy effective models \cite{lopesdossantos_graphene_2007, lopes_dos_santos_continuum_2012,bistritzer_moire_2011, koshino_interlayer_2015, koshino_maximally_2018, guinea2019continuum, rost_map_2019, lee2019theory, koshino2020effective, CarrFang_PRR_2019, fang2019angle, kang2022pseudo, vafek2022continuum} were developed, by which the
most interesting physics in TBG systems at the magic angle – flat bands, has been predicted theoretically. In the widely used Bistritzer-MacDonald (BM) model \cite{bistritzer_moire_2011}, the 
moir\'e potential is expanded to leading order by Fourier Transformation and a general model Hamiltonian for arbitrary 
twist angles can be constructed. With such model, B\&M predict that flat bands emerge when twist angle is around 1.1$^\circ$ which is called ``magic angle". The ``magic" happens because 
the low energy band is so flat that the velocity of electrons vanishes. In these flat band systems, the electron-electron interactions dominate the band structure and could result in various exotic correlated phenomena. More accurate calculation based on density functional theory (DFT) \cite{uchida_atomic_2014, lucignano_crucial_2019,cantele_structural_2020,PhysRevB.105.125127} can also be performed using large scale parallelized DFT code integrated with van der Waals functional to catch the details of the band structure. However, this kind of computation is very time consuming and hard to implement for systems with small twist angles.

Although BM model provides a clear picture of flat bands near the first magic angle and is easy to implement numerically, the model itself ignores microscopic details such as atomic relaxation \cite{nnt2017, carr2018relax, uchida_atomic_2014, lucignano_crucial_2019, gupta2019straintronics, fleischmann2019perfect, cantele_structural_2020, gargiulo2017structural,guinea2019continuum, anglieD6,leconte_relaxation_2022} in the moir\'e scale which can explain the insulating phase of MATBG at $\pm 4$ filling. For many 2D materials, tight binding (TB) models are often used to describe electronic band structures. Compared with BM model, tight binding scheme can easily take atomic relaxation into consideration by resetting the coordinates of carbon atoms and the related hopping parameters can be determined by fitting to small scale DFT calculation results. Full tight binding calculation \cite{morell2010flat, tra2010nano, moon2012energy, moon_optical_2013,kangprx2018} in the framework of Slater-Koster theory \cite{slater_simplified_1954} is performed and provides reliable results. Several \textit{ab initio} TB models \cite{fang2016electronic, fang2019angle, pathak2022accurate, davydov2022construction, kang2022pseudo, vafek2022continuum} have been carefully designed to make the band structures of TBG much closer to DFT results. However, full TB model will generate a huge Hamiltonian matrix when twist angle is small which makes further calculations hard to perform.

In this manuscript, we present a well designed numerical scheme named Truncated Atomic Plane Wave (TAPW) method to project the full tight binding Hamiltonian onto the truncated atomic plane waves. TAPW shows very accurate band structures compared with those retrieved from full TB Hamiltonian. The key point of our new method is to combine the advantages of both the BM model and full TB model. First, like full TB model, the atomic plane wave basis will be constructed with full information of the real atomic positions. Next, like the treatment in BM model, for small enough twist angles, the atomic valley degree of freedom can still be treated as a good quantum number and then the complete Hilbert space of full TB model can be truncated into two groups of atomic plane waves, where only the wave vectors close enough to the graphene $\bK/\bK'$ points are included. After the basis truncation, the total dimension of the Hamiltonian can be reduced significantly from several tens of thousand to only several hundreds. Our scheme has the following advantages compared with existing models:
\begin{enumerate}
    \item It saves computing resources compared with a full TB scheme but shares the same accuracy of low energy bands.
    \item The parameters of the TB model can be carefully designed, using a Slater Koster scheme \cite{morell2010flat, tra2010nano, moon2012energy, moon_optical_2013,kangprx2018} or an \textit{ab initio} TB scheme \cite{fang2016electronic, fang2019angle, pathak2022accurate, davydov2022construction, kang2022pseudo, vafek2022continuum}.
    \item The whole workflow is simple and straightforward without manually expanding BM model to higher orders \cite{CarrFang_PRR_2019, guinea2019continuum, koshino2020effective, kang2022pseudo, vafek2022continuum}, or fitting parameters \cite{CarrFang_PRR_2019, po_faithful_2019}.
    \item Follow up studies, such as Hartree Fock (HF), constrained Random Phase Approximation (cRPA) or dynamic mean field (DMFT) calculations can be carried out based on our method.
    \item The relaxation effect and symmetry constraint can be considered appropriately by setting real space coordinates of carbon atoms like other TB models.
\end{enumerate}
Based on this kind
of numerical strategy, we develop an open source \texttt{Python} package hosted on GitHub \footnote{\href{https://github.com/zybbigpy/TAPW}{GitHub Repository Link for \texttt{TAPW}}}. We also extend our method to describe the 
moir\'e phonons and determine the screened Coulomb interaction of Magic Angle Twisted Bilayer Graphene (MATBG) using constrained Random Phase Approximation (cRPA).

\section{From BM model to Truncated Atomic Plane Wave Method}

\subsection{Geometry of TBG}
We define $\mathbf{a}_1 = a(\sqrt{3}/2, -1/2), \mathbf{a}_2 = a(\sqrt{3}/2, 1/2)$ as the lattice vectors for the atomic structure of monolayer graphene and 
$\mathbf{b}_1 = 2\pi/a(\sqrt{3}/3, -1), \mathbf{b}_2 = 2\pi/a(\sqrt{3}/3, 1)$ as their corresponding reciprocal lattice vectors. $a=0.246 \, \text{nm}$ is the graphene lattice constant. The geometry of TBG can be
defined by rotating two different layers of AA-stacking bilayer graphene around the AA-stacking point. After anti--clockwisely rotating the layer (1) by $+\theta/2$  and the layer (2) by $-\theta/2$, the lattice vectors should be $\mathbf{a}_i^{(1)} = D\left(\frac{\theta}{2}\right) \mathbf{a}_i, \mathbf{a}_i^{(2)} = D\left(-\frac{\theta}{2}\right) \mathbf{a}_i$ and the reciprocal lattice vectors are $\mathbf{b}_i^{(1)} = D\left(\frac{\theta}{2}\right) \mathbf{b}_i,
\mathbf{b}_i^{(2)} = D\left(-\frac{\theta}{2}\right) \mathbf{b}_i$, where $D(\theta)$ is a 2D rotation matrix. 

\begin{figure}
\includegraphics[width=0.48\textwidth]{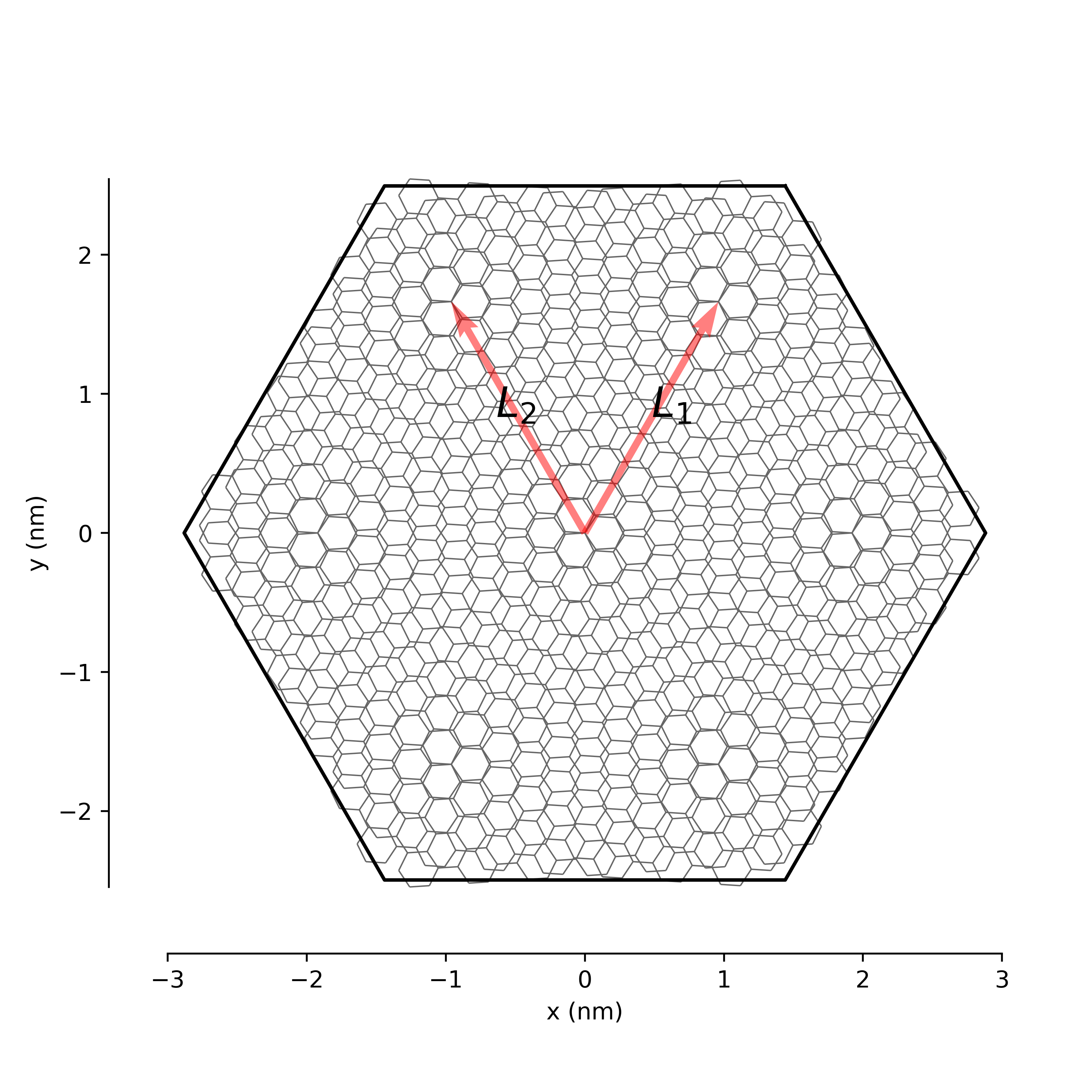}
\caption{Schematic diagram for the real space twist using the geometry stated in the main context when $N=4$, i.e. $\theta=7.341^\circ$. $\mathbf{L}_1, \mathbf{L}_2$ are moir\'e unit lattice vectors. 
Corresponding $\bk$ space diagram is shown in Fig.\ref{fig:ksampling}.
}
\label{fig:realspace}
\end{figure}

TBG system will have a commensurate structure \cite{lopesdossantos_graphene_2007, mele2010commensuration, lopes_dos_santos_continuum_2012} if the twist angle obeys:
\begin{equation}
    \theta = \arcsin \left(\frac{\sqrt{3}(2N+1)}{6N^2+6N+2}\right),
\end{equation}
where $N$ is an integer. As illustrated in Fig.\ref{fig:realspace}, the unit lattice vectors for the moir\'e super cell are:
\begin{equation}
    \begin{aligned}
        \mathbf{L}_1 &= D\left(\frac{\theta}{2}\right)\left[-N \mathbf{a}_1 + (2N+1)\mathbf{a}_2\right],\\
        \mathbf{L} _2 &= D\left(\frac{\theta}{2}\right)\left[-(2N+1)\mathbf{a}_1 + (N+1)\mathbf{a}_2\right].
    \end{aligned}
\end{equation}
The corresponding moir\'e reciprocal lattice vectors can be chosen as:
$ \bG_1 = \mathbf{b}_1^{(1)} - \mathbf{b}_1^{(2)},\bG_2 = \mathbf{b}_2^{(1)} - \mathbf{b}_2^{(2)}.$ The system has a $D_3$ point group symmetry \footnote{Systems with a $D_6$ point group symmetry are also provided in our \texttt{Python} Packages. }.
\subsection{BM Model}
A widely used way to deal with the band structure of TBG at small twist angle is BM model \cite{lopesdossantos_graphene_2007, lopes_dos_santos_continuum_2012, bistritzer_moire_2011, koshino_interlayer_2015, koshino_maximally_2018}.  In BM model, the Hamiltonian of valley $\xi= \pm1$ written in real space is:
\begin{equation}
    H_\xi(\br)=\left[\begin{array}{cc}
    h^{(1)}(\br) & U(\br) \\
    U^{\dagger}(\br) & h^{(2)}(\br)
    \end{array}\right],
    \label{eq:contham}
\end{equation}
where $h(\br)$ represents for intralayer part of the Hamiltonian and $U(\br)$ is the large scale moir\'e potential. The analysis for BM can be done in $\bk$-space. For a two-layer system like TBG, the Bloch basis for each layer (or atomic Bloch basis) can be defined as:
\begin{equation}
    \label{eq:atomic_bloch}
    \begin{aligned}
        \psi_{\bk,\alpha}^{(1)}(\br) &= \frac{1}{\sqrt{N}} \sum_{\bRu_\alpha} \e^{\ri \bk \cdot \bRb_\alpha} \phi_{p_z}(\br-\bRb_\alpha), \\
        \psi_{\bp,\beta}^{(2)} (\br)&= \frac{1}{\sqrt{N}} \sum_{\bRu_\beta} \e^{\ri \bp\cdot \bRu_\beta} \phi_{p_z}(\br-\bRu_\beta),
    \end{aligned}
\end{equation}
where $N$ is the number of atomic unit cell in each layer, and $\bR_\alpha,\bR_\beta$ is the concrete position of A/B carbon atoms, (1)/(2) is a notation for different layers.

Under this kind of basis function, $h(\br)$ is naturally diagonalized with $\bk$ and can be approximated as a Dirac equation when 
$\bk$ or $\bp$ is near atomic $\bK, \bK'$ points. However, $U(\br)$ will couple different $\bk$ and $\bp$ satisfying the 
following condition \cite{koshino_interlayer_2015}:
\begin{equation}
    \label{eq:conserve}
    \bk=\bp-m_1 \bG_1-m_2 \bG_2,
\end{equation}
$m_1,m_2$ are two integers. BM approximates $U(\br)$ by keeping three leading Fourier components:
\begin{equation}
    \begin{aligned}
    U(\br)=&\left[\begin{array}{cc}
    U_{\mathrm{A}_{1} \mathrm{A}_{2}} & U_{\mathrm{A}_{1} \mathrm{B}_{2}} \\
    U_{\mathrm{B}_{1} \mathrm{A}_{2}} & U_{\mathrm{B}_{1} \mathrm{B}_{2}}
    \end{array}\right] \\
    =&\left[\begin{array}{cc}
    u & u^{\prime} \\
    u^{\prime} & u
    \end{array}\right]+\left[\begin{array}{cc}
    u & u^{\prime} \omega^{\xi} \\
    u^{\prime} \omega^{-\xi} & u
    \end{array}\right] \e^{i \xi (-\mathbf{G}_{1})\cdot \mathbf{r}}\\ 
    &+\left[\begin{array}{cc}
    u & u^{\prime} \omega^{-\xi} \\
    u^{\prime} \omega^{\xi} & u
    \end{array}\right] \e^{i \xi\mathbf{G}_{2} \cdot \mathbf{r}},
    \end{aligned}
\end{equation}
where $\omega = \mathrm{e}^{(\ri 2\pi/3)}$, and we adopt $u=0.08581 \,\mathrm{eV}, u' = 0.1032\, \mathrm{eV}$ after considering corrugation effect. Parameters are retrieved from our new method, see Appx.\ref{app:table}.

\subsection{Truncated Atomic Plane Wave Method: Theoretical Formalism and Numerical Strategies}
The BM model is a good approximation for low energy physics of TBG at the first magic angle. However, band structures solved from BM model are not reliable any more when twist angle gets smaller where relaxation effect plays an important role \cite{nnt2017, gargiulo2017structural, carr2018relax, CarrFang_PRR_2019, anglieD6, leconte_relaxation_2022}. Parameters like $u, u'$ in the BM model are highly sensitive to the structure of TBG and more Fourier components of $U(\br)$ should be considered. To develop a more accurate description for the low energy physics of TBG, we can still follow the main idea and keep the key approximation of BM model:

\begin{enumerate}
    \item Expand the Hamiltonian using atomic Bloch function defined in Eq.(\ref{eq:atomic_bloch}), see ref.~\cite{bistritzer_moire_2011, koshino_interlayer_2015, koshino_maximally_2018, guinea2019continuum, koshino2020effective}.
    \item Intervalley tunneling process can be safely ignored at small twist angle, so the system has an approximate $U_v(1)$ symmetry, see ref.~\cite{lopesdossantos_graphene_2007, bistritzer_moire_2011, CarrFang_PRR_2019, fang2019angle}.
    \item Since the low energy physics mainly comes from atomic
     $\bK$ or $\bK'$ points, only a group of plane waves close to Dirac points should be taken into consideration, see ref.~\cite{bistritzer_moire_2011, koshino_maximally_2018, guinea2019continuum, koshino2020effective, kang2022pseudo, vafek2022continuum}.
\end{enumerate}

These observations from the BM model studies first inspire us to directly expand the Hamiltonian of TBG utilizing atomic Bloch basis defined in Eq.(\ref{eq:atomic_bloch}). When the system is commensurate, the atomic Bloch basis could be modified to match the moir\'e superlattice. More concretely, the following substitution should work
\begin{equation}
    \begin{aligned}
        N   &= N_{\mathrm{m}}N_\mathrm{a},\,\, \bk = \bkbar + \bG_n,
    \end{aligned}
\end{equation}
where $N_{\mathrm{m}}$ is the number of moir\'e lattices and
$N_{\mathrm{a}} = M/4$,
$M$ is the number of all atoms in a moir\'e superlattice.
$\bkbar$ is defined in the first moir\'e B.Z. and $\bG_n$ is the moir\'e reciprocal lattice vector.
These substitution transforms the atomic Bloch function, which is Eq.(\ref{eq:atomic_bloch}), to
\begin{equation}
\label{eq:atomic_basis}
\ket{\psi_{\alpha n}(\bkbar)} = \frac{1}{\sqrt{N_\mathrm{m} N_\mathrm{a}}}\sum_{\I, i} \e^{ \ri(\bkbar+\bG_n)\bR_{\I i \alpha}} \ket{\phi_{p_z}(\br-\bR_{\I i \alpha})},
\end{equation}
where $\bR_{\I i \alpha} = \bL_{\I} + \btau_{i \alpha}$ for short and $\bL_{\I}$ is the lattice vector of the moir\'e unit cell. $\alpha=$ A$_1$, B$_1$, A$_2$, B$_2$ denotes the sublattice, $\btau_{i\alpha}$ is the displacement of atom $\alpha$ in
the $i$-th atomic cell with respect to the I-th moir\'e cell. The basis wavefunction, which we call \textit{atomic plane
wave basis}, can be viewed as a Bloch summation of atomic $p_z$ orbitals, and for unrelaxed TBG, it is naturally normalized.

After using atomic plane wave basis defined in Eq.(\ref{eq:atomic_basis}) 
to expand the Hamiltonian and further taking advantage of Eq.(\ref{eq:conserve}), we can write down the matrix element $H_{\alpha n, \beta m}$  of the Hamiltonian:

\begin{equation}
\begin{aligned}
  &\mel{\psi_{\alpha n}(\bkbar)}{\hat{H}}{\psi_{\beta m}(\bkbar)} \\
=& \frac{1}{N_\mathrm{m} N_\mathrm{a}} \sum_{\I \J,ij} t(\bR_{\I i \alpha}-\bR_{\J j \beta})
\e^{-\ri(\bkbar+\bG_n)\bR_{\I i \alpha}} \e^{\ri(\bkbar+\bG_m)\bR_{\J j \beta}} \\
=&\frac{1}{N_\mathrm{m} N_\mathrm{a}} \sum_{\I \J,ij} \e^{-\ri \bG_n \btau_{i\alpha}} \e^{-\ri \bkbar (\bL_\I -\bL_\J+\btau_{i\alpha}-\btau_{j\beta})} \\
&\times t(\bL_\I -\bL_\J+\btau_{i\alpha}-\btau_{j\beta}) \e^{\ri \bG_m \btau_{j\beta}}\\
=& \frac{1}{N_\mathrm{m} N_\mathrm{a}} \sum_{\I ,ij} \e^{-\ri \bG_n \btau_{i\alpha}} \e^{-\ri \bkbar (\bar{\btau}_{i\alpha, j\beta})} 
t(\bar{\btau}_{i\alpha, j\beta}) \e^{\ri \bG_m \btau_{j\beta}} \\
=&\sum_{ij} \left(\frac{\e^{-\ri \bG_n \btau_{i\alpha}}}{\sqrt{N_{\mathrm{a}}}}\right) \cdot\left[\e^{-\ri \bkbar (\bar{\btau}_{i\alpha, j\beta})} 
t(\bar{\btau}_{i\alpha, j\beta})\right]\cdot \left( \frac{\e^{\ri \bG_m \btau_{j\beta}}}{\sqrt{N_{\mathrm{a}}}}\right). \\
\label{eq:ele}
\end{aligned}
\end{equation}
$\I,\J$ are moir\'e lattice indices and $i\alpha, j\beta$ are position indices for carbon atoms in the moir\'e superlattice. $\alpha,\beta$ runs over $[\mathrm{A}_1,\mathrm{B}_1, \mathrm{A}_2, \mathrm{B}_2]$. $\bar{\btau}_{i\alpha,j\beta}$ is the distance between atom $i\alpha$ and atom $j\beta$. $t(\bar{\btau}_{i\alpha, j\beta})$ is the hopping integral under tight binding approximation. Furthermore, the matrix form of the Hamiltonian, which we 
denote as $\bH^{\text{TAPW}}$, can be written in a more compact way by inspecting the last step of Eq.(\ref{eq:ele})
\begin{equation}
\begin{aligned}
\bH^{\text{TAPW}} &= \sum_{\alpha \beta} \mathbf{X}^\dagger_{\alpha} \mathbf{T}_{\alpha\beta} \mathbf{X}_{\beta}\\
&= \mathbf{X}^\dagger \mathbf{T} \mathbf{X},
\label{eq:Hmat}
\end{aligned}
\end{equation}
where the corresponding matrix elements are
$(\mathbf{X_\alpha})_{n,i}=\e^{\ri\bG_n\btau_{i\alpha}}/\sqrt{N_\mathrm{a}}$, $(\mathbf{T}_{1\alpha \beta})_{i, j} = \e^{-\ri \bkbar (\bar{\btau}_{i\alpha, j\beta})}$,
$(\mathbf{T}_{2\alpha \beta})_{i, j} =t(\bar{\btau}_{i\alpha, j\beta})$, 
$\mathbf{T}= \mathbf{T}_1 * \mathbf{T}_2$, and ``*" is an element-wise product. 

\begin{figure}
\includegraphics[width=0.52\textwidth]{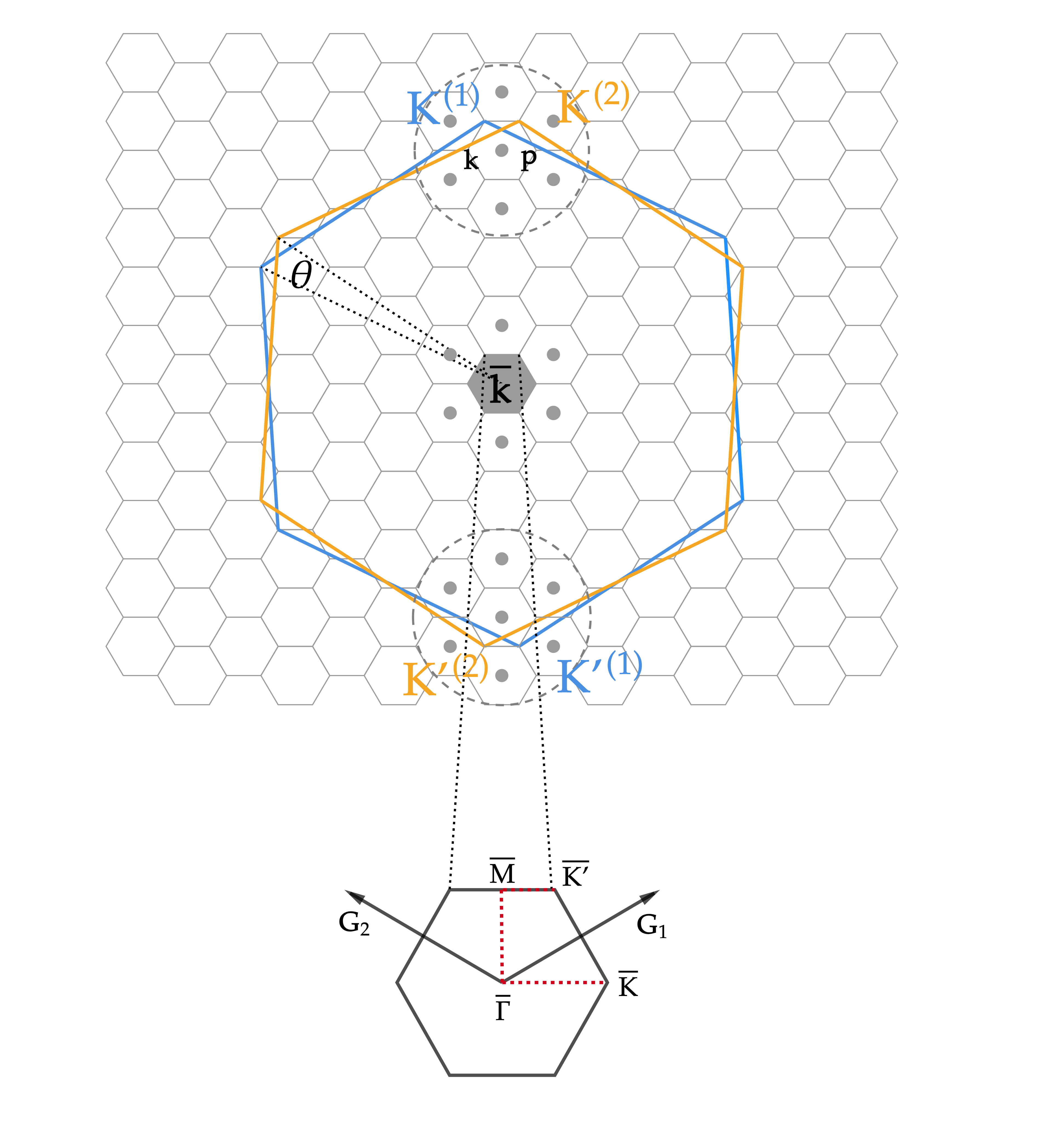}
\caption{Schematic diagram for rotation in the  $\bk$ space for $N=4$, i.e. $\theta = 7.341^\circ$. We construct $\bG$-list in two circled areas (centered around $\bK,\bK'$) for two different valleys when calculating the band structure for moir\'e electrons and 
construct a $\bG$ list centered at $\Gamma$ point when calculating the moir\'e phonon bands. Two big hexagons represent for the reciprocal lattice of two graphene sheets and the smaller grey one is moir\'e reciprocal lattice. High symmetry points of the moir\'e reciprocal lattice are denoted as $\bar{\mathbf{\Gamma}},\bar{\bK},\bar{\mathbf{M}},\bar{\bK'}$. $\bG_1, \bG_2$
are moir\'e reciprocal lattice vector. The small grey dots represent the plane waves used to expand the TB Hamiltonian with the first shell conserved. The band structures of TBG are calculated along high symmetry path (denoted in a red dashed line).
}
\label{fig:ksampling}
\end{figure}

In Eq.(\ref{eq:Hmat}), $\mathbf{T}$ is exactly the full tight binding matrix and $\mathbf{X}$ is
a plane wave projector under continuum approximation $\e^{\ri\bG_n\btau_{i\alpha}} = \e^{\ri\bG\cdot\mathbf{r}}\delta(\mathbf{r}-\btau_{i\alpha}) \approx \e^{\ri\bG\cdot\mathbf{r}} $. When $\bG$ runs over all the moir\'e reciprocal lattice vectors in the graphene first B.Z., Eq.(\ref{eq:Hmat}) is a unitary transformation and it will restore the full TB model. Following the spirit of the BM model, we do \textit{truncation} on the $\bG$ vectors within certain fixed distance away from graphene $\bK/\bK'$ points when calculating the band structure of TBG at a small twist angle, as illustrated in Fig.\ref{fig:ksampling}. We want to further emphasize,
our method is intrinsically equivalent to the generalized BM model because we start from the same basis function (atomic Bloch function) to expand the Hamiltonian. However, our method requires the system to be commensurate, then Eq.(\ref{eq:atomic_basis}) can be established. 
The advantage of our new method is that TAPW directly projects the full TB matrix on a series of plane waves without assuming the tunneling amplitude $t$ between different layers is a smooth function and then manually performing Fourier expansion \cite{bistritzer_moire_2011, koshino_interlayer_2015, koshino_maximally_2018, CarrFang_PRR_2019, guinea2019continuum, koshino2020effective, kang2022pseudo, vafek2022continuum}. Interestingly, our TAPW method presents an exact mapping between the full TB Hamiltonian and generalized continuum model, which is a simple realization of  ref.~\cite{rost_map_2019}.

In our numerics, the hopping integral $t(\bar{\btau}_{i\alpha, j\beta})$ is determined by Slater-Koster (SK) formula \cite{moon_optical_2013,koshino_maximally_2018}:
\begin{equation}
t(\br) = -V_\pi\left(1-\frac{r_z^2}{r^2}\right)-V_\sigma\frac{r_z^2}{r^2},
\label{eq:SK}
\end{equation}
with $r_z = \br \cdot \mathbf{e}_z, V_\pi = V_\pi^0 \e^{-(r-a_0)/r_0}, V_\sigma = V_\sigma^0 \e^{-(r-d_0)/r_0}.$
where $d_0= 0.335\,\text{nm}$ is the average interlayer spacing, $a_0=a/\sqrt{3}$ is the nearest neighbour distance, $r_0 = 0.184a$ is the characteristic length of the hopping strength, the hopping amplitudes are set as $V_\pi^0 = -2.7\,\text{eV}$ and  $V_\sigma^0=0.48\, \text{eV}$. It is worthwhile to mention that the hopping parameter $t(\bar{\btau}_{i\alpha, j\beta})$ can be replaced by a more accurate SK formula \cite{haddadi2020moire} or environment adapted \textit{ab initio} results \cite{fang2016electronic, pathak2022accurate, kang2022pseudo, vafek2022continuum}, see Appx.\ref{app:more_tb}.

For a better understanding of the superiority of our TAPW method compared with the full TB model, we then introduce the numerical details in our realization. Note that the size of $\bG$--list is $N_G$ and the number of atoms in the moir\'e superlattice is $M= 4\times N_{\mathrm{a}}$. The dimension of the Hamiltonian matrix $\mathbf{H}$ is $4N_G\times4N_G$, the tight binding matrix $\mathbf{T}$ is $M\times M$ and the plane wave projection matrix $\mathbf{X}$ is $4N_G\times M$. Our TAPW method projects the sparse TB matrix $\mathbf{T}$ into a much smaller subspace. Typically for MATBG ($\theta = 1.085^\circ$, $M=11164$), we can restore electronic band structure near Fermi level ($\pm 0.6$ eV) perfectly compared with full TB result using only 244 plane waves per valley, i.e., $N_G=61$. The TB Hamiltonian for each $\bk$ point is downfolded from $11164\times 11164$ to $244\times244$ per valley. We transform the computational complexity of diagonalizing a huge sparse matrix into the multiplication of sparse matrices together with diagonalizing a much smaller dense one. The latter operation saves a huge amount of computational power and preserves low energy electronic band structure (see detailed discussion in Sec.\ref{sec:numerical}). The sparse matrix operations are boosted by \texttt{SciPy} \cite{2020SciPy-NMeth}. It is worth pointing out the setup of $\mathbf{X}, \mathbf{T}_2$ is only once during the whole computational process because they are not $\bkbar$ dependent.

The construction of plane wave projection matrix $\mathbf{X}$ is fast and the corresponding computational complexity is $\mathcal{O}(M\times N_G)$. However, setting up transfer integral matrix $\mathbf{T}_2$ and hopping phase matrix $\mathbf{T}_1$ is very time consuming because we have to determine the neighbours of a specific carbon atom in such a large system with more than 10,000 atoms. The brute force searching scheme has a complexity of $\mathcal{O}(M^2)$ and it fails full TB model in dealing with a much smaller twist angle.

\begin{figure}
\includegraphics[width=0.50\textwidth]{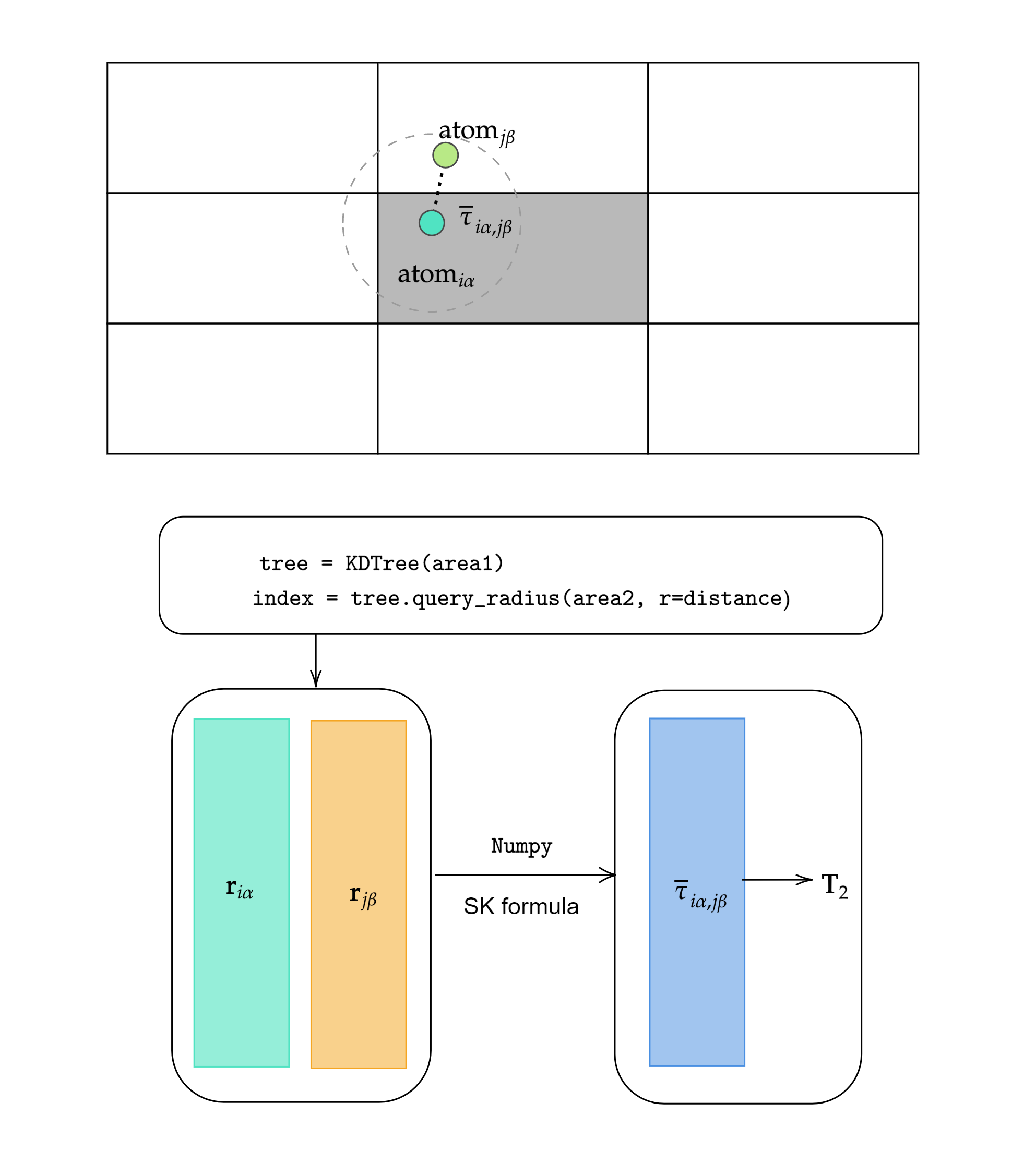}
\caption{We design an efficient numerical scheme to build the 
hopping matrix $\mathbf{T}_2$. The grey block represents for the moir\'e super cell and we set up a 3 $\times$ 3 super cell to perform a $k$-d tree search. The grey block is labeled as $\texttt{area2}$ and the larger one as $\texttt{area1}$.  The searching process is performed using $k$-d tree algorithm integrated in $\texttt{sklearn}$ \cite{sklearn_api} which returns the neighbour pair indices $(i\alpha, j\beta)$. 
The 3D coordinates for neighbour pairs $\mathbf{r}_{i\alpha}-\mathbf{r}_{j\beta}$ are stored in \texttt{NumPy ndarray} denoted with different color blocks. The computation for hopping parameter $\bar{\bm{\tau}}_{i\alpha, j\beta}(\mathbf{r}_{i\alpha}-\mathbf{r}_{j\beta})$ is finished in $\mathcal{O}(1)$ time taking advantage of the built in vectorization mechanism of $\texttt{NumPy}$. Then the hopping matrix $\mathbf{T}_2$ can be constructed with neighbour pair indices and $\bar{\bm{\tau}}$ array.
}
\label{fig:kdtree} 
\end{figure}

Another strategy we adopt is using $k$-d tree \cite{kdtree} to optimize the searching scheme of determining neighbours. We utilize the fact there are only carbon atoms in the TBG system and the hopping process can mainly happen between one specific carbon atom and another from the nearest moir\'e unit cell. Thus, we construct a $3\times3$ super cell as the searching space. Such super cell may consist of millions of carbon atoms but it only consumes a small amount of memory space to store 3D coordinates. As demonstrated in Fig.\ref{fig:kdtree}, a highly efficient $k$-d tree searching scheme is applied which reduces the related computational complexity to $\mathcal{O}(M\log M)$ and $t(\bar{\btau}_{i\alpha, j\beta})$ is calculated in $\mathcal{O}(1)$ time enhanced by the vectorization characteristic of \texttt{NumPy} \cite{harris2020array}.

\begin{figure}
\includegraphics[width=0.48\textwidth]{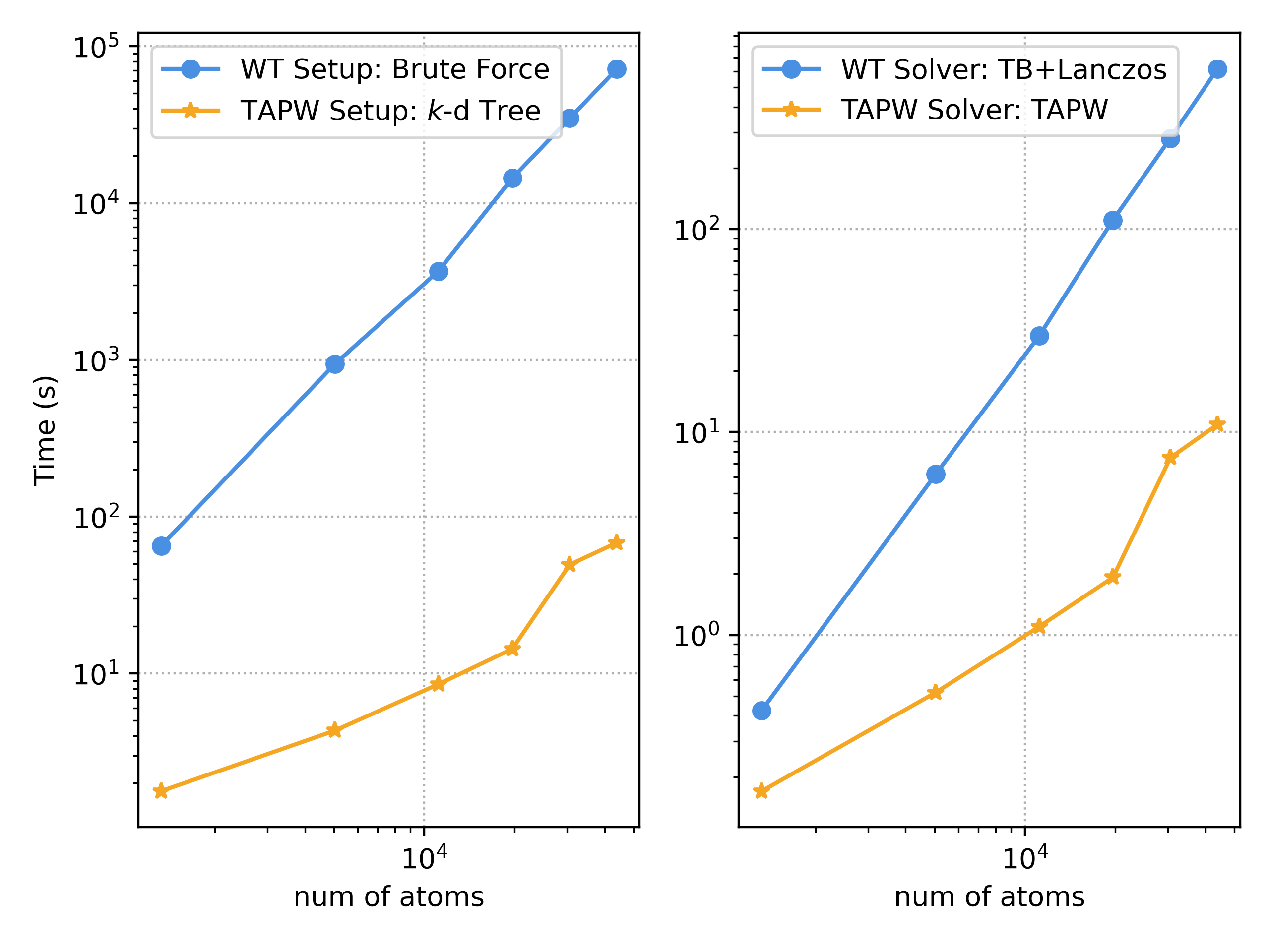}
\caption{\label{fig:benchmark}Computational efficiency comparison between full tight binding scheme and 
truncated atomic plane wave scheme. Tight binding solver for twisted bilayer graphene is a new feature of \texttt{WannierTools (WT)} , 
a popular \texttt{Fortran} routine to solve tight binding Hamiltonian and related topological properties. \texttt{TAPW} is a submodule 
of our \texttt{Python} package. The left panel plots the time consumed to set up the tight binding kernel versus number of atoms in the moir\'e superlattice. \texttt{WT} uses a brute force searching scheme to build SK tight binding kernel while \texttt{TAPW} uses a $k$-d tree searching 
scheme. The right panel plots the time consumed to diagonalize the Hamiltonian per $\mathbf{k}$ point versus the number of atoms in the moir\'e superlattice.  \texttt{WT} utilizes the power of \texttt{ARPACK} (Lanczos algorithm) to calculate the eigenvalues of the sparse tight binding Hamiltonian and \texttt{TAPW} uses the proposed projection algorithm. The benchmark is performed serially on an 8-core, 16-thread Intel Xeon W-3223 processor with \texttt{WannierTools V2.6.2}.
}
\end{figure}

\subsection{Corrugation and Relaxation}
In TAPW, the information of atomic position in the moir\'e scale is encoded through tight binding description by definition. The deviation of TBG system from a rigid structure can be considered naturally by setting the 3D coordinates of carbon atoms. As observed in the DFT calculation \cite{uchida_atomic_2014, lucignano_crucial_2019, cantele_structural_2020} 
and molecular dynamics simulations \cite{gargiulo2017structural,guinea2019continuum, anglieD6, leconte_relaxation_2022} for TBG near the first magic angle, the two graphene sheets are not totally flat and there exists some fluctuation in the real space. The $z$ direction displacement can help separate flat bands away from remote bands near the first magic angle and further stabilize the insulating phase of MATBG at $\pm 4$ filling \cite{lucignano_crucial_2019}. This kind of corrugation effect can be easily simulated in our TAPW method by adding displacements along $z$ direction for carbon atoms \cite{uchida_atomic_2014, koshino_maximally_2018},

\begin{equation}
    \begin{aligned}
        d^{(1)} &= \frac{1}{2}d_0 + d_1 \sum_{n=1,2,3} \cos \left( \bG_n \cdot \btau \right), \\
        d^{(2)} &= \frac{1}{2}d_0 - d_1 \sum_{n=1,2,3} \cos \left( \bG_n \cdot \btau \right),
    \end{aligned}
\end{equation}
where $d_0=3.43$ \r{A} is average interlayer distance for TBG, $d_1 =0.278$ \r{A} is obtained from looking at the difference of the interlayer distance 
between AA--stacking bilayer and AB--stacking bilayer. $\bG_3 = - (\bG_1+\bG_2)$ is the third smallest moir\'e reciprocal lattice vector. $\bm{\tau}$ is the atomic position of the carbon atom in the moir\'e unit cell. 

Moreover, the local geometry of rigid TBG can be classified into three different regions: AA-stacking area, AB/BA-stacking area and saddle point (SP) area. As pointed out in DFT study \cite{gargiulo2017structural}, AB/BA-stacking area and SP area are more energetically favorable compared with AA-stacking area, but the in-plane strain field can also compete with such kind of interlayer energy minimization. It is also evidenced in STM experiments \cite{Choi2019}, when the twist angle is small, TBG undergoes a self-organized lattice reconstruction to shrink AA-stacking area and expand AB-stacking area which forms a triangular
lattice in the moir\'e scale. These microscopic in-plane distortions can be well
captured in TAPW by reconstructing the Bloch function \cite{koshino2020effective}:
\begin{equation}
    \begin{aligned}
        \psi_{\bk,\alpha}^{(1)}(\br) &= \frac{1}{\sqrt{N}} \sum_{\bRu_\alpha} \e^{\ri \bk \cdot \bRb_\alpha} \phi_{p_z}(\br-\bRb_\alpha-\bm{u}(\bRb_\alpha)), \\
        \psi_{\bp,\beta}^{(2)}(\br) &= \frac{1}{\sqrt{N}} \sum_{\bRu_\beta} \e^{\ri \bp\cdot \bRu_\beta} \phi_{p_z}(\br-\bRu_\beta-\bm{u}(\bRu_\beta)).
    \end{aligned}
\end{equation}
$\bm{u}$ is an abstract displacement vector field which slowly varies in the atomic scale with a moir\'e periodicity. We can still follow the proposed procedure to perform calculation and take valley as a good quantum number when the twist angle is small. The Hamiltonian can be solved in an elegant way by just resetting hopping matrix $\mathbf{T}_2$ with the relaxed coordinates of the carbon atoms. The full atomic relaxation can be performed using classical molecular dynamics with the rigid structure as a start point. The electronic band structures of relaxed MATBG computed using TAPW method are summarized in Appx.\ref{app:more_tb}.

\subsection{Numerical Results and Comparison}\label{sec:numerical}

\begin{figure*}
\includegraphics[width=0.88\textwidth]{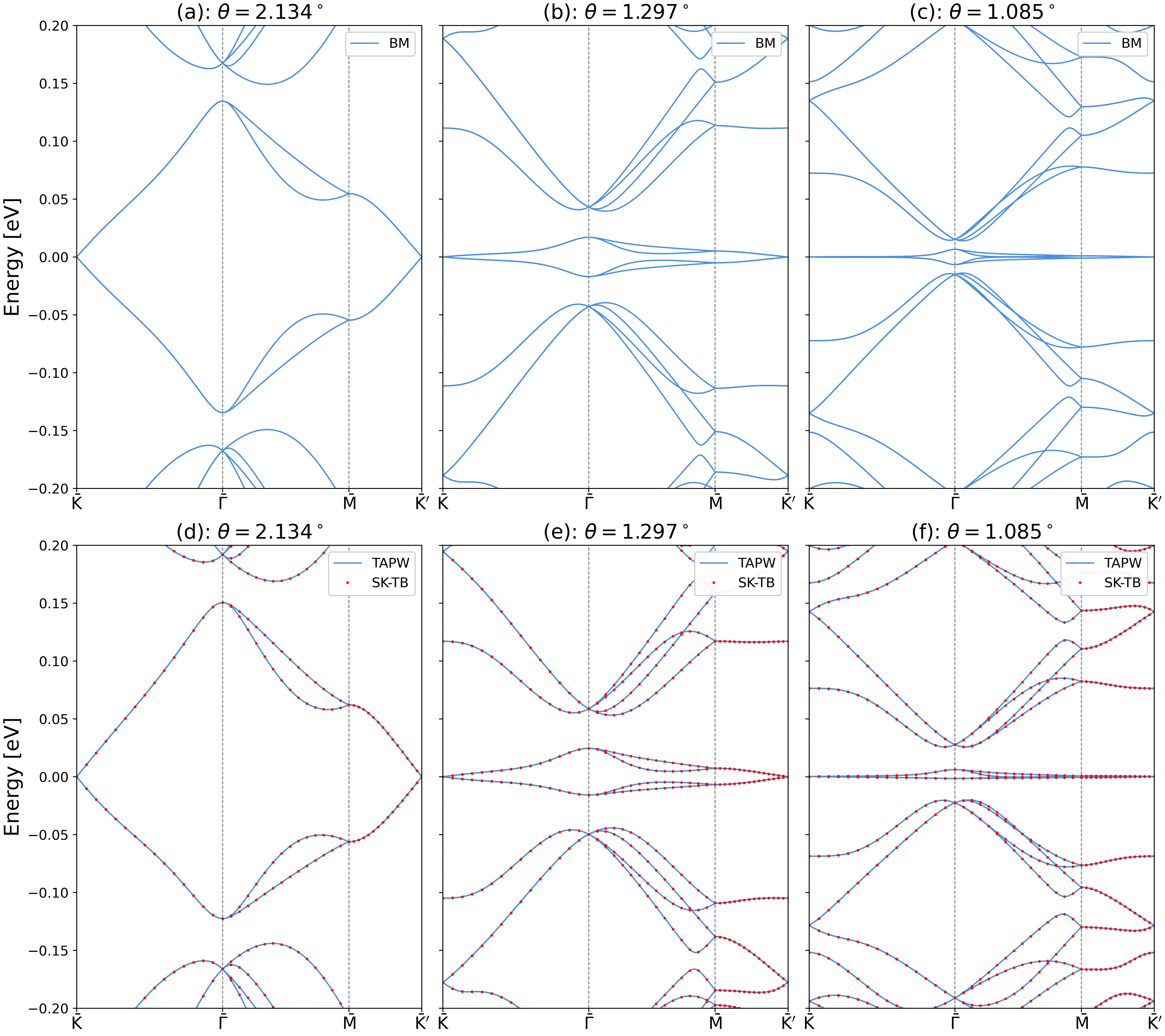}
\caption{\label{fig:bandcomp} Electronic Band structures of TBG along high symmetry points $\bar{\bK}-\bar{\mathbf{\Gamma}}-\bar{\mathbf{M}}-\bar{\bK'}$ at different twist
angles using different methods when corrugation effect is taken into consideration. (a)-(c): Band structures computed using BM model for $\theta=2.134^\circ, 1.297^\circ, 1.085^\circ$, respectively.
(d)-(f): Band structures computed using full TB (SK-parameterization) and TAPW method for $\theta=2.134^\circ, 1.297^\circ, 1.085^\circ$, respectively. The full TB reference results are denoted in red dots while the TAPW results are in blue lines. TB results show a particle hole asymmetry compared with BM model.}
\end{figure*}
\begin{table*}
\caption{\label{tab:comparison}Comparison between different kinds of Band Calculation Methods in TBG system.}
\begin{ruledtabular}
\begin{tabular}{lccc}
    Methods &  Full TB model         & BM model   &TAPW  method\\  \hline
    Basis Size                 & number of atoms & 4$\times$ number of $\bG$ vectors & $4\times$ number of $\bG$  vectors       \\ 
    Computational Cost         & High            & Low                & Medium                   \\
    Computational Accuracy      & High                        & Medium             & High                     \\ 
    Corrugation and Relaxation   & Set Coordinates    & Set $u, u'$    & Set Coordinates \\ 
    Incommensurate System        &\XSolidBrush              & $\checkmark$     &\XSolidBrush      \\ 
    Easy to Use                 &\XSolidBrush              & $\checkmark$     & $\checkmark$   
\end{tabular}
\end{ruledtabular}
\end{table*}

Based on the theoretical formalism of TAPW, we developed a robust \texttt{Python} package to perform calculation. Numerical schemes including full TB method, TAPW method and BM model are all realized. Their characteristics are briefly summarized in Table.\ref{tab:comparison}.

We take \texttt{WannierTools} (\texttt{WT}) \cite{wu2018wanniertools} as a benchmark for the full TB calculation. In TB calculation, the whole process can be divided into two parts: setting up the hopping matrix $\mathbf{T}_2$ or hopping integral file \texttt{hr\_dat} and diagonalizing the TB Hamiltonian. Our numerical strategy takes several seconds to set up the hopping integral matrix $\mathbf{T}_2$ at the first magic angle ($\theta=1.085^\circ$) while \texttt{WT} takes more than one hour to build the hopping integral file \texttt{hr\_dat}. \texttt{TAPW} presents a huge advantage when the twist angle gets even smaller, as clarified on the left panel of Fig.\ref{fig:benchmark}.
For band structure calculations, we diagonalize the projected matrix instead of struggling with a huge sparse matrix like \texttt{WT} (\texttt{WT} diagonlizes the sparse TB Hamiltonian using Lanczos algorithm integrated in \texttt{ARPACK} \cite{lehoucq1998arpack}, this feature is also realized in our \texttt{Python} package.) 
Our method saves a large amount of time and reproduces low energy band structures perfectly in a wide range of twist angles, as shown in the right panel of Fig.\ref{fig:benchmark} and Fig.\ref{fig:bandcomp}, respectively.

We denote $\mathbf{v}_t$ to be the $t-$th eigen vector (column vector) of the full TB matrix ($t$ is labeled from Fermi level) and $\bar{\mathbf{v}}_t$ to be the corresponding one of the projected TB matrix. Then we take $\eta = |\mathbf{v}_{t}^{\dagger}\cdot \mathbf{X}\cdot\bar{\mathbf{v}}_t|$ as a criteria to evaluate the performance of TAPW basis. The better it will be for TAPW basis if $\eta$ is closer to 1 ($\mathbf{X}\cdot\bar{\mathbf{v}}$ restores the full tight binding eigenvector). At the first magic angle, we take 244 atomic plane waves per valley (488 bands in total) to expand the full TB matrix and find that $\eta > 99\%$ for eigen vectors of four flat bands at the $\bar{\Gamma}$ point.
Now one can be convinced that TAPW method outputs high quality eigen wavefunctions.

Compared with BM model, our method generates a similar matrix structure to describe the 
Hamiltonian if the same $\bG$--list is used. In BM model, people do Fourier analysis for  $U_{\alpha,\beta}(\bk, \br)$, as commonly analyzed in ref.~\cite{koshino_maximally_2018}:
\begin{equation}
    \begin{aligned}
        U_{\alpha,\beta}(\mathbf{k}, \mathbf{r}) =& \sum_{m_1, m_2} \tilde{U}_{\alpha,\beta}(m_1 \mathbf{b}_1 + m_2\mathbf{b_2}+\mathbf{k})\\
                                         &\times\exp[\ri(m_1 \mathbf{b}_1 + m_2\mathbf{b_2})\cdot\bm{\delta}_{\alpha,\beta}]\\
                                         &\times \exp[\ri(m_1 \mathbf{G}_1 + m_2\mathbf{G_2})\cdot\mathbf{r}],
    \end{aligned}
\end{equation}
where,
\begin{equation}
    \tilde{U}_{\alpha,\beta}(\mathbf{q}) = -\frac{1}{S_0}\int t[\mathbf{R}+d(\mathbf{R}-\bm{\delta}_{\alpha,\beta}) \mathbf{e}_z]\mathrm{e}^{-\ri \mathbf{q}\cdot {\mathbf{R}}} \mathrm{d}\mathbf{R}.
\end{equation}
BM model adopts two important assumptions for the description of moir\'e potential. 
\begin{enumerate}
    \item $\tilde{U}(\mathbf{q})$ decays in $q \approx 1/r_0$ because the transfer integral $t$ is determined by Slater Koster formula in Eq.(\ref{eq:SK}) which exponentially decays in $R \approx r_0$. That's the reason why BM model only conserves three largest Fourier components: $(m_1, m_2)=(0, 0), (-1, 0), (0, 1)$ for $\bK$ valley.
    \item When performing $\mathbf{k}$ sampling in the area that is close to $\mathbf{K}$, BM model always assumes $U(\mathbf{k}, \mathbf{r}) = U({\mathbf{K}, \mathbf{r}})$. It is a good approximation that $\tilde{U}(|\mathbf{k}|)\approx \tilde{U}(|\mathbf{K}|)$ at the first magic angle when $|\mathbf{k}-\mathbf{K}|\leq|\mathbf{G}|\approx |\mathbf{K}|$ \cite{bistritzer_moire_2011}, because $|\mathbf{K}|\approx 30 |\mathbf{G}|\gg|\bG|$. 
\end{enumerate}

In common practice, $\mathbf{k}$ sampling is performed in the area that $|\mathbf{k}-\mathbf{K}|\leq 4|\mathbf{G}|$ \cite{koshino_maximally_2018} when calculating the band structure of MATBG. When $|\mathbf{k}-\mathbf{K}| \leq |\mathbf{G}|$, it brings us the accurate flat band picture which is the original BM model \cite{bistritzer_moire_2011}. When $|\mathbf{k}-\mathbf{K}|>|\mathbf{G}|$, these assumptions are rough and result in more deviation on the bands away from Fermi surface compared with the full TB results, as shown in Fig.5 of ref. \cite{moon_optical_2013}. Additionally, atomic relaxation at small twist angle will make high order $\bG$ components of moir\'e potential more significant than the rigid one \cite{CarrFang_PRR_2019, guinea2019continuum,CarrFang_PRR_2019,fang2019angle, kang2022pseudo, vafek2022continuum}. 

\begin{figure*}
\includegraphics[width=0.88\textwidth]{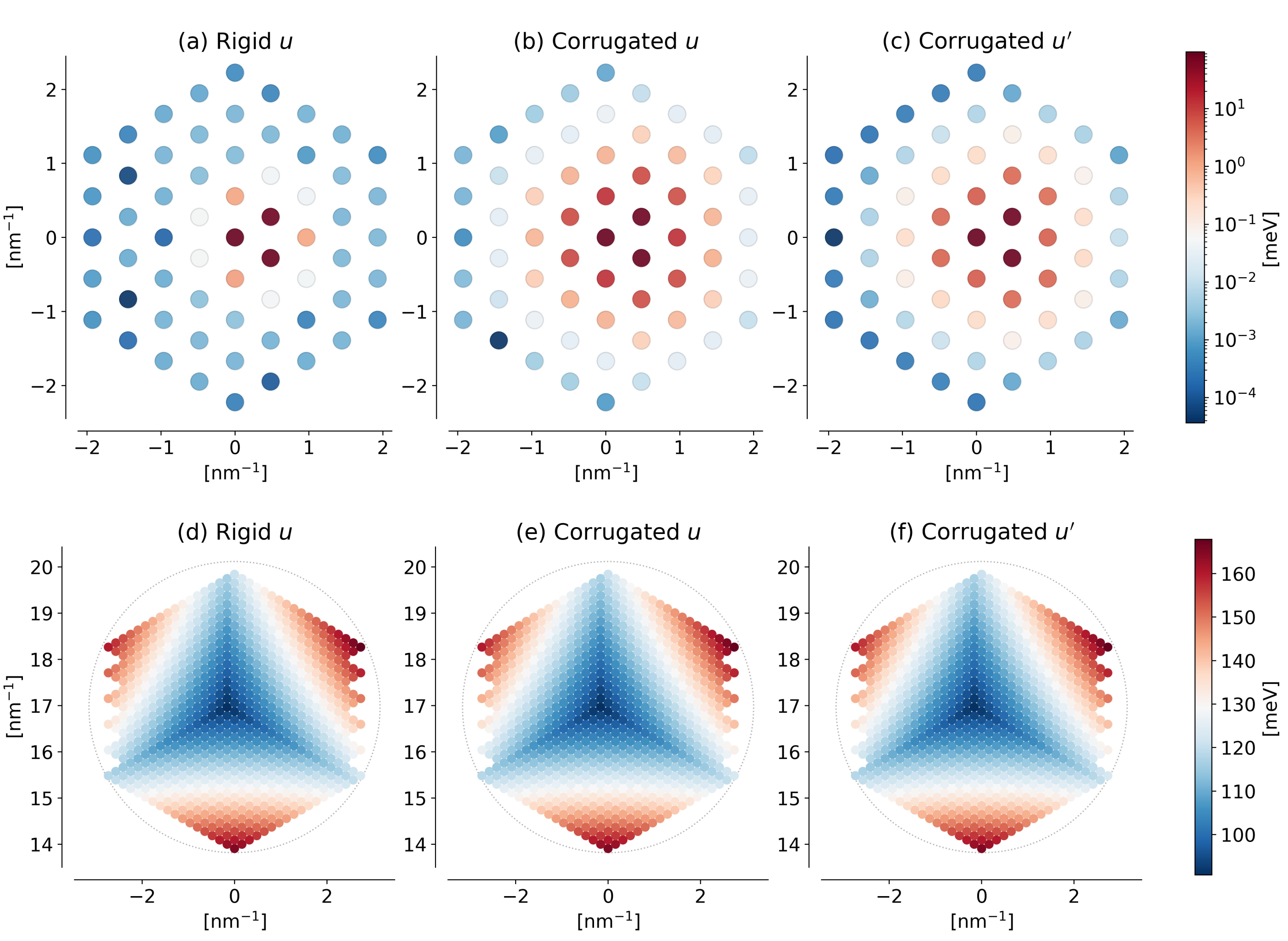}
\caption{\label{fig:moirepot} Moir\'e potential $U_{\alpha,\beta}(\bk,\bG)$ calculated using TAPW method when $\theta=1.085^\circ$. (a)-(c): $\bG$ dependence of $U$ when $\bk$ is fixed at $\Gamma+\bG_0$, $\bG_0=30(\bG_1+\bG_2)$. The largest $\bG$ component is located at $\bG_0$, $\bG_0-\bG_2$, $\bG_0+\bG_1$ as the ones from BM. As shown, corrugation effect brings more high order $\bG$ components into effect. (d)-(f): $\bk$ dependence of $U$ when the largest $\bG$ component is reserved which BM totally smears out. (a),(d): Rigid structure as input, $u=|U_{\mathrm{AA}}|=|U_{\mathrm{AB}}|$. (b),(e): Corrugated structure as input, $u=|U_{\mathrm{AA}}|$. (c),(f) Corrugated structure as input, $u'=|U_{\mathrm{AB}}|$.}
\end{figure*}

Above analysis suggests a more accurate description for the moir\'e potential is needed.
Different from existing studies which expand BM manually to higher orders \cite{koshino2020effective, CarrFang_PRR_2019, fang2019angle,guinea2019continuum, kang2022pseudo, vafek2022continuum} and integrate a $k$-linear term in the interlayer coupling \cite{koshino2020effective, CarrFang_PRR_2019, fang2019angle}, our new method automatically expands the moir\'e potential into a tensor form $U_{\alpha\bG_i, \beta \bG_j}(\bkbar)$, where $\bkbar+\bG_i=\bk\approx \mathbf{K}$. As demonstrated in the first row of Fig.\ref{fig:moirepot} where we fix $\bk = \Gamma+\mathbf{G}_0 \approx 1/2(\mathbf{K}^{(1)}+\mathbf{K}^{(2)})$, corrugation in MATBG brings high order $\mathbf{G}$ components into effect and creates a difference between $u$ and $u'$. In the second row of Fig.\ref{fig:moirepot} ,
we see $U$ does rely on $\bk$ by conserving the largest $\mathbf{G}$ component of $U$ at different $\bk$ points in the area $|\bk-\mathbf{K}|<5|\bG|$. These corrections on BM provide us a clear picture for electron and hole asymmetry in the flat bands.

\section{Applications}

\subsection{Low Frequency Moir\'e Phonons}
\begin{figure*}
\includegraphics[width=0.88\textwidth]{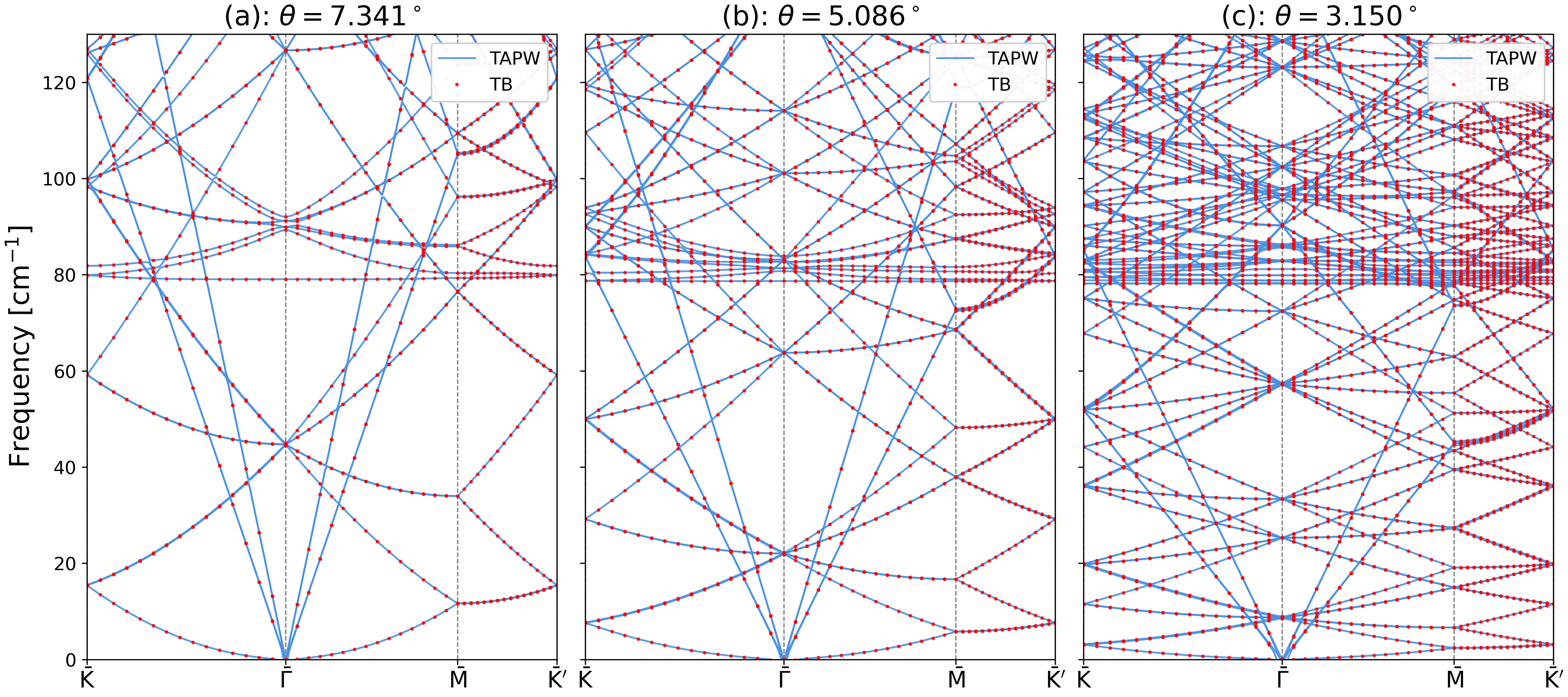}
\caption{\label{fig:bandphonon} Low frequency phonon band structures of TBG along high symmetry points $\bar{\bK}-\bar{\mathbf{\Gamma}}-\bar{\mathbf{M}}-\bar{\bK'}$ at different twist angles, (a): $\theta=7.341^\circ$, (b): $\theta =5.086^\circ$, (c): $\theta = 3.150^\circ$. The full tight binding results are denoted in red dots while TAPW bands are in blue lines. The dispersionless layer breathing (LB) mode is located at $\omega \approx 80\, \mathrm{cm}^{-1}$.}
\end{figure*}
Moir\'e phonons 
\cite{koshino2019moire, suri2021chiral,maity2022chiral, lu2022low} also received great attention as novel collective phenomena are observed in twisted bilayer graphene \cite{gadelha2021localization} and twisted MoS$_2$ \cite{lin2018moire, quan2021phonon} using Raman spectra technique. It is natural to generalize our TAPW method for moir\'e phonons by mapping tight binding Hamiltonian for electrons to dynamic matrix for lattice vibrations. In this section, we provide a rigorous derivation of TAPW method for moir\'e phonons and study the low frequency moir\'e phonons for TBG systems as an example.

The equation of motion for the phonon field $u_\nu(\bR_{\J j \beta})$ for TBG under harmonic approximation can be written as:
\begin{equation}
    \frac{1}{M_c} \sum_{\J j \beta \nu} \Phi_{\mu \nu} (\bR_{\I i\alpha}-
    \bR_{\J j \beta}) u_\nu(\bR_{\J j \beta}) = \omega^2 u_\mu(\bR_{\I i\alpha}).
\end{equation}
$M_c$ is the mass for the carbon atoms and $\Phi_{\mu \nu} (\bR_{\I i\alpha}-
\bR_{\J j \beta})$ is the force constant between two carbon atoms and $\mu, \nu$ represent for the Cartesian coordinates. After Fourier Transforming the phonon field taking advantage of the moir\'e periodicity and Born–von Karman boundary condition, we get:
\begin{equation}
    u_\mu (\bR_{\I i\alpha}) = \frac{1}{\sqrt{N_\mathrm{m}}} \sum_{\bar{\bq}\in \text{m.B.Z.}} \tilde{u}_{i\alpha\mu} (\bar{\bq}) \e^{\ri \bar{\bq}\cdot\bR_{\I i\alpha}},
\end{equation}
which leads to the full tight binding description for moir\'e phonons:
\begin{equation}
    \sum_{j\beta\nu} D_{i\alpha \mu, j\beta\nu} (\bqbar) \tilde{u}_{j\beta\nu} (\bqbar) = \omega^2(\bqbar) \tilde{u}_{i\alpha\mu}(\bqbar),
\end{equation}
and the dynamic matrix $D_{i\alpha \mu, j\beta\nu} (\bqbar) $ defined in the moir\'e B.Z. is:
\begin{equation}
    D_{i\alpha \mu, j\beta\nu} (\bqbar) = \frac{1}{M_c} \sum_{\bR_{\J}} \Phi_{\mu \nu} (\mathbf{0}+\bm{\tau}_{i\alpha}-
    \bR_{\J j \beta}) \e^{\ri \bqbar\cdot(\bR_{\J}+\bm{\tau}_{j\beta}-\bm{\tau}_{i\alpha})}.
\end{equation}
For moir\'e systems like TBG, we can always expand phonon fields using the periodicity of single layer graphene:
\begin{equation}
\begin{aligned}
    u_\mu (\bR_{\I i\alpha}) &= \frac{1}{\sqrt{N}} \sum_{\bq\in\text{B.Z.}} \bar{u}_{\alpha\mu} (\bq) \e^{\ri \bq\cdot\bR_{\I i \alpha}}\\
    &= \frac{1}{\sqrt{N_{\mathrm{m}}N_\mathrm{a}}} \sum_{\bqbar, \bG_n} \bar{u}_{\alpha\mu} (\bqbar+\bG_n) \e^{\ri( \bqbar \cdot\bR_{\I i \alpha}+\bG_m\cdot \bm{\tau}_{i\alpha})}.
\end{aligned}
\label{eq:phonon}
\end{equation}
Substitute Eq.(\ref{eq:phonon}) to the equation of motion for the phonon field and use the definition of dynamic matrix $D_{i\alpha \mu, j\beta\nu} (\bqbar) $, we arrive:
\begin{equation}
    \sum_{\beta\nu m} \bar{D}_{\alpha\mu n, \beta\nu m} (\bqbar) \bar{u}_{\beta\nu} (\bqbar+\bG_m) = \omega^2(\bqbar) \bar{u}_{\alpha\mu}(\bqbar+\bG_n),
\end{equation}
where the projected dynamic matrix $\bar{D}_{\alpha\mu n, \beta\nu m}(\bqbar)$ is defined as:
\begin{equation}
\bar{D}_{\alpha\mu n, \beta\nu m}(\bqbar) = \frac{1}{N_{\mathrm{a}}}\sum_{ij} \e^{-\ri \bG_n\cdot \bm{\tau}_{i\alpha}}D_{i\alpha \mu, j\beta\nu} (\bqbar)\e^{\ri \bG_m\cdot \bm{\tau}_{j\beta}}.
\end{equation}
The corresponding matrix form reads:
\begin{equation}
\begin{aligned}
\bar{\mathbf{D}}^{\text{TAPW}} &=  \sum_{\alpha \beta}\mathbf{X}^\dagger_{\alpha} \mathbf{D}_{\alpha\beta} \mathbf{X}_{\beta}\\
&= \mathbf{X}^\dagger \mathbf{D} \mathbf{X},
\label{eq:Dmat}
\end{aligned}
\end{equation}
and $\alpha,\beta$ now are joint indices for Cartesian coordinates index, A/B sublattice index and layer index. For low frequency phonon bands, we construct a truncated $\bG$--list centered at $\Gamma$ point instead of $\bK$ or $\bK'$ point in the problem of electrons. 

\begin{figure*}
\includegraphics[width=0.88\textwidth]{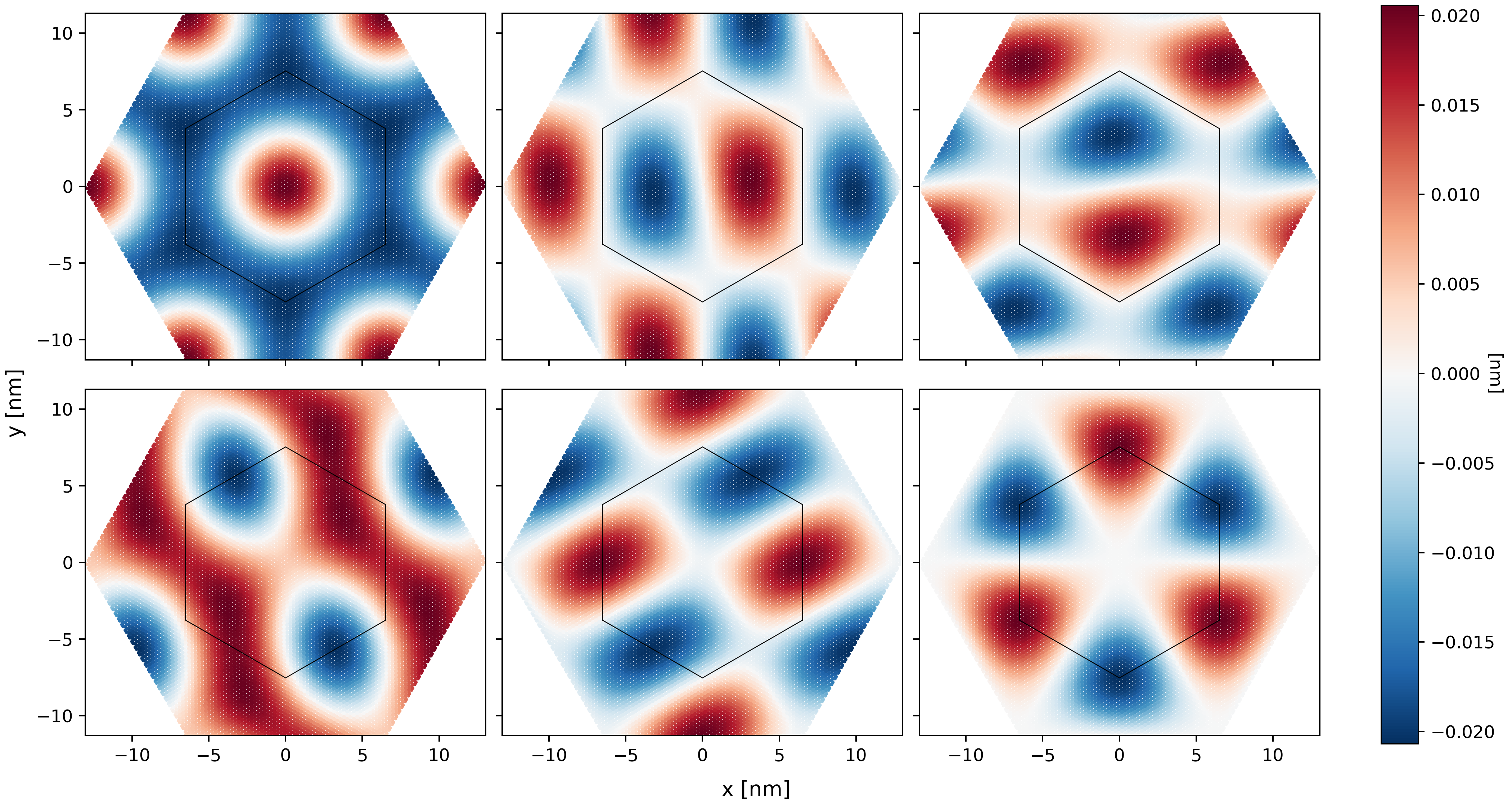}
\caption{\label{fig:optical_mode} Low frequency optical phonon modes solved at $\bar{\Gamma}$ point of MATBG using TAPW method. The black hexagon marks the Wigner–Seitz cell of TBG. We find  all-symmetric-type, dipolar-type, quadrupolar-type and octupolar-type
out-of-plane vibrations as in ref.\cite{liu2021phonons}.}
\end{figure*}

We use frozen phonon method to compute the force constants $\Phi_{\mu \nu}(\mathbf{R}_{\mathrm{I}i\alpha}-\mathbf{R}_{\mathrm{J}j\beta})$ from the relaxed structure:
\begin{equation}
    \Phi_{\mu \nu}(\mathbf{R}_{\mathrm{I}i\alpha}-\mathbf{R}_{\mathrm{J}j\beta})=\frac{\partial^{2} U}{\partial \mathbf{R}_{\mu \mathrm{I}i\alpha} \partial \mathbf{R}_{\nu \mathrm{J}j\beta}}=-\frac{\partial\mathbf{F}_{\nu \mathrm{J}j\beta}}{\partial \mathbf{R}_{\mu \mathrm{I}i\alpha}}.
\end{equation}
where $U = U_{\mathrm{bonded}} + U_{\mathrm{non-bonded}} $ is the potential energy consisting of the bonded intra-layer interactions and non-bonded van der Waals inter-layer interactions. The bonded interactions can be modelled by the Dreiding force fields \cite{mayo1990dreiding}, and the non-bonded van der Waals interactions are modelled by an exponential-6 form \cite{mayo1990dreiding, pascal2010quantum}. Details are described in Appx.\ref{app:ff}. Before computing the force constants, the lattice was relaxed to optimize the geometry by performing conjugate gradient (CG) algorithm embedded in \texttt{LAMMPS} \cite{LAMMPS}. With relaxed lattice, the force constants can be approximated by finite displacement method \cite{phonopy} with a small displacement of 0.01 \AA. 

The full dynamic matrix $\mathbf{{D}}$ can then be constructed like full TB Hamiltonian for electrons:
\begin{equation}
    \mathbf{{D}} = \mathbf{T}_1 * \mathbf{\Phi},
\end{equation}
which is an element-wise product with $\mathbf{T}_1$ being hopping phase matrix. As shown in Fig.\ref{fig:bandphonon}, low frequency phonon bands solved at a series of twist angles of TBG using our TAPW method show a perfect consistency with the ones from direcly diagonalizing dynamic matrix $\mathbf{{D}}$. Furthermore, polarization vector $\tilde{u}_{i \alpha \mu} (\bqbar)$ can be restored using 
eigen vectors $\bar{u}_{\alpha\mu}(\bqbar+\bG_n)$ of $\mathbf{\bar{D}}^{\text{TAPW}}$:
\begin{equation}
    \tilde{u}_{i \alpha \mu} (\bqbar) = \frac{1}{\sqrt{N_\mathrm{a}}} \sum_{n} \bar{u}_{\alpha \mu} (\bqbar+\bG_n) \mathrm{e}^{\ri \bG_n\cdot \btau_{i\alpha}},
\end{equation}
which has a simple matrix form $\tilde{\mathbf{u}} = \mathbf{X}\bar{\mathbf{u}}$. We plot out-of-plane vibrations resulting from low frequency optical phonon modes of MATBG at $\bar{\Gamma}$ point (around 0-30 cm$^{-1}$) using our TAPW method in Fig.\ref{fig:optical_mode}. The result is consistent with the ones calculated in ref.~\cite{liu2021phonons} where the authors directly solve a huge dynamic matrix (33492 $\times$ 33492). In our calculation, we use 732 atomic plane waves centered at $\Gamma$ point to expand the full dynamic matrix and the projected dynamic matrix is reduced to $732\times732$. Then we can finish the whole computation on a laptop. (The number of atomic plane waves is $61\times12=732$ when $N_G=61$.)
Based on above numerical experiments, we believe that TAPW method can not only provide accurate band structures but also detailed eigen wavefunctions for moir\'e phonons.

\begin{figure*}
\includegraphics[width=\textwidth]{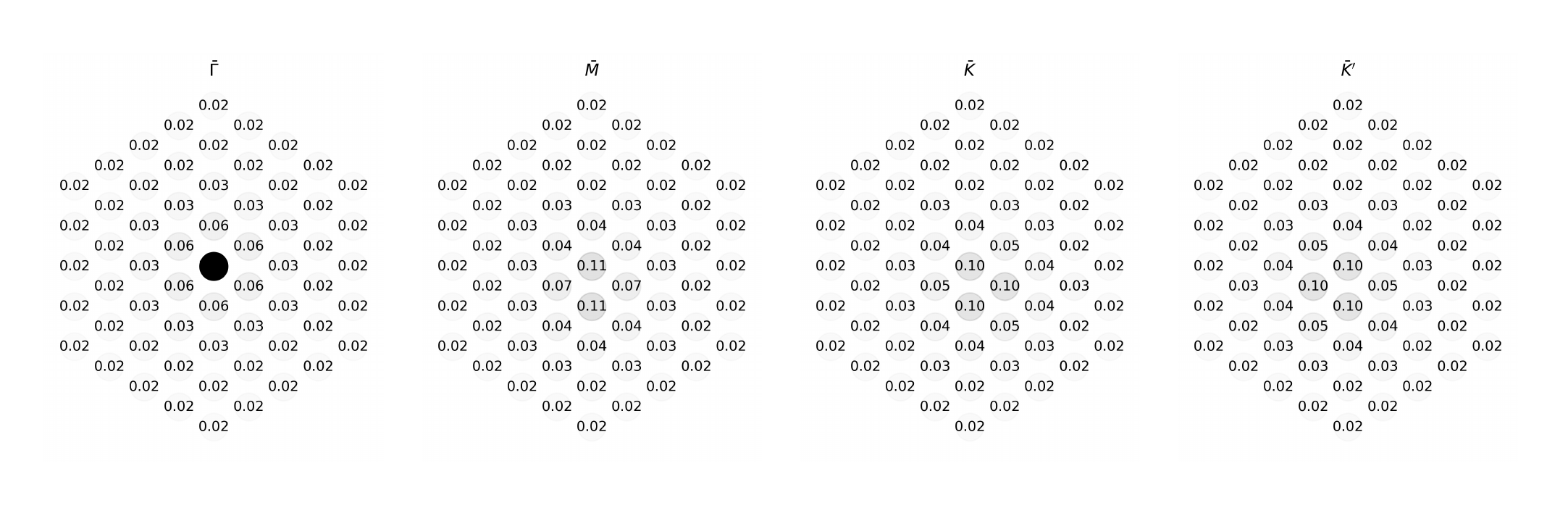}
\caption{\label{fig:ueff}Screened Coulomb interaction calculated when twist angle $\theta=1.085^\circ$ on high symmetry points. The $\bG$-list 
is constructed in a hexagonal shape. We set $\kappa = 0.005\textup{\AA}^{-1}, \varepsilon = 5$. The value of $U_{\bQ,\bQ'}(\bkbar)/U_{\bQ_0, \bQ_0}(\bar{\Gamma})$ is
plotted on the $\bQ$ component, $U_{\bQ_0, \bQ_0}(\bar{\Gamma})=0.198\mathrm{eV}$. }
\end{figure*}

\subsection{Constrained Random Phase Approximation}

Several unrestricted Hartree Fock calculations \cite{Cocemasov2013, zhang_correlated_2020, liu_theories_2021, zhang2021correlated, kwan2021kekule, wagner2022global, haoshi2022} have been performed to study the competing orders in the flat band system of MATBG. However, most of the studies use a manually designed single gated or double gated form of Coulomb interactions. A more accurate screened Coulomb interaction should be computed if virtual particle hole exchange from remote bands is taken into consideration.

As discussed in Sec.\ref{sec:numerical}, our TAPW method generates high resolution band structures, not only flat bands but also remote bands, compared with the ones solved from BM model. In this section, we introduce a reliable calculation scheme to determine the screened Coulomb interaction form \cite{goodwin2019attractive, vanhala2020constrained, pizarro2019internal} in MATBG using the technique of constrained Random Phase Approximation (cRPA) \cite{aryasetiawan_crpa_2004,aryasetiawan_2006}.

Based on previous band structure calculation, it is obvious that the flat bands of MATBG are well separated from those high energy bands. cRPA allows people to study the screened Coulomb interaction in this kind of 
narrow band system. In cRPA, 
the single particle Hilbert space is divided into two parts, which we 
call the $r$ and $d$ subspace. The $d$ space contains low energy narrow bands while the 
$r$ space hosts high energy bands.
The total polarization $\Pi$ of the system can be separated into two parts,  $\Pi_d$ is the polarization within 
the narrow bands and $\Pi_r$ is the rest of the polarization:
\begin{equation}
\Pi = \Pi_d + \Pi_r.
\end{equation}
The totally screened interaction $W_r$ can be calculated in the following way:
\begin{equation}
W_r = \frac{U}{1-U\Pi_r},
\end{equation}
where $U$ is bare Coulomb interaction.

For MATBG, the Coulomb interaction term can be 
written directly as:

\begin{equation}
H_{\mathrm{I}} = \frac{1}{2S}\sum_{\bk\bk'\bq<\Lambda}\sum_{\alpha,\alpha'}
U(\bq)c^\dagger_{\bk+\bq,\alpha}c^\dagger_{\bk'-\bq,\alpha'} c_{\bk',\alpha'}c_{\bk,\alpha},
\end{equation}
with $U(\bq)= e^2/(\varepsilon\varepsilon_0\sqrt{q^2+\kappa^2})$ as bare Coulomb interaction. $\kappa$ is the inverse screening length and $\varepsilon$ is background dielectric constant. Here, the interaction Hamiltonian is written in the atomic Bloch basis, and $\bk,\bk',\bq$ can be truncated near 
Dirac points (denoted as $\Lambda$). Rewrite the equation 
in the extended atomic plane wave basis, we get:

\begin{equation}
    \begin{aligned}
        &H_{\mathrm{I}} = \frac{1}{2S}\sum_{ss'}\sum_{\bkbar\bkbar'\bar{\bq}} 
\sum_{\bG_n\bG_{n'}\mathbf{Q}_m <\Lambda} \sum_{\alpha \alpha'} U(\bar{\bq}+\mathbf{Q}_m)\\
&c^\dagger_{s,\bkbar+\bar{\bq}+\bG_n+\mathbf{Q}_m,\alpha} c^\dagger_{s',\bkbar'-\bar{\bq}+\bG_{n'}-\mathbf{Q}_m,\alpha'}
c_{s',\bkbar'+\bG_{n'},\alpha'}
c_{s,\bkbar+\bG_n,\alpha},
    \end{aligned}
\end{equation}
where $s$ is the index for spin and valley (we consider there's no spin-valley flipping),
$\bkbar,\bkbar', \bar{\bq}$ are defined in the moir\'e B.Z. and $\bG_n,\bG_{n'},\mathbf{Q}_{m}$ are all 
moir\'e reciprocal lattice vector. $S$ is the size of the real space, which equals to $N_\bk \Omega$ ($\Omega$
is the size of moir\'e unit cell.)

In cRPA, the non--dynamic screened interaction for flat bands reads:
\begin{equation}
U_{\mathrm{screened}}(\bar{\bq}) = U(\bar{\bq})(1-U(\bar{\bq})\Pi_0(\bar{\bq}))^{-1}.
\end{equation}
The matrix form of bare interaction $U(\bar{\bq})$ is defined by
\begin{equation}
U(\bar{\bq})_{m,m'} = U(\bar{\bq}+\mathbf{Q}_m) \delta_{m,m'},
\end{equation}
and polarization tensor $\Pi_{0}(\bar{\bq})_{\mathbf{Q}_m \mathbf{Q}_{m'}}$ is:
\begin{equation}
\begin{aligned}
	\Pi_0(\bq)_{m,m'} = &\frac{2}{S}\sum_{\bkbar}\sum_{\bG_n\bG_{n'}}\sum_{\ell \ell'\alpha\alpha'}\\
	&\braket{\bG_n \alpha}{E_{\ell} (\bkbar)}\braket{E_{\ell'}(\bkbar+\bar{\bq})}{\bG_n+\mathbf{Q}_m \alpha}\\
	\times&\braket{E_{\ell}(\bkbar)}{\bG_{n'}\alpha'} \braket{\bG_{n'}+\mathbf{Q}_{m'}\alpha'}{ E_{\ell'}(\bkbar+\bar{\bq})}\\
\times&\frac{n({E_\ell(\bkbar))}-n(E_{\ell'}(\bkbar+\bar{\bq}))}{E_\ell(\bkbar)-E_{\ell'}(\bkbar+\bar{\bq})}.
\end{aligned}
\end{equation}
The summation of band index $\ell$ should exclude the ones for flat bands. 

We determine the screened Coulomb interaction $U_{\mathrm{screened}}(\bar{\bq})_{\mathbf{Q}\mathbf{Q}'}$ for a corrugated TBG structure when twist angle $\theta=1.085^\circ$. We set up the single particle wavefunctions using our TAPW method on a $6\times6$ $\bk$-mesh and then perform cRPA calculation to retrieve screened Coulomb interaction. The numerical results are presented in Fig.\ref{fig:ueff}. Single gated or double gated Coulomb interaction can be replaced by $U_{\mathrm{screened}}(\bar{\bq})_{\mathbf{Q}\mathbf{Q}'}$  to improve the credibility of HF results. 

\section{Summary and Outlook}

As a summary, we present TAPW, a carefully optimized numerical scheme to downfold the full ``tight binding" Hamiltonian for TBG to a low energy effective model using a series of truncated atomic plane waves. Our method shows a perfect consistency with the low energy bands solved from full tight binding Hamiltonian, not only for moir\'e electrons but also for moir\'e phonons. This kind of low cost projection can be generalized to other twisted moir\'e systems if 
credible tight binding Hamiltonian is constructed to describe electrons or faithful dynamic matrix for phonons. For example, parameterized tight binding Hamiltonian for twisted transition metal dichalcogenides (TMDCs) \cite{fangTBTMD, vitale2021flat} has been proposed recently. Lattice dynamics of several important 2D materials can be simulated using semi-classical molecular dynamics integrated with modern force fields \cite{hBNff, hBNGff}. In this paper, we
further visualize moir\'e potential of MATBG by projecting interlayer interaction $U$ to a series of atomic plane waves which clarifies the importance of high order $\bG$ components and related $\bk$ dependence in the band structure at small twist angles. 

TAPW provides a systematical way to study moir\'e electrons and phonons in a single particle manner. By freezing lattice vibrations induced by specific phonon mode, TAPW can build low energy effective model containing electron phonon coupling \cite{choi2018strong, choi2021dichotomy}. As pointed out in ref.~\cite{angeli2019valley, angeli2020jahn, blason_distortion_2022}, the iTO phonon-induced local Kekul\'e distortion may correspond to fruitful phase diagrams in MATBG and this kind of distortion has been observed in a recent nano Raman experiment \cite{gadelha2021localization}.  

Another feature of TAPW is the model itself is intrinsically equivalent to the generalized BM model. The projected Hamiltonian shares the same structure which makes further computation easy to implement. We carry out cRPA calculation to determine screened Coulomb interaction of TBG at the first magic angle. This kind of screened Coulomb interaction can be taken as a lower bound to replace commonly used single gated or double gated Coulomb interaction in Hartree Fock calculation.
\begin{acknowledgments}
We thank fruitful discussion with Tianyu Qiao and Hao Shi. W.M. acknowledges support via the UC Santa Barbara NSF Quantum Foundry funded via the Q-AMASE-i program under award DMR-1906325. D.P. acknowledges support from the Hetao Shenzhen/Hong Kong Innovation and Technology Cooperation (HZQB-KCZYB-2020083). X.D. acknowledges financial support from the Hong Kong Research Grants Council
(No. 16309020). 

\end{acknowledgments}
\appendix
\begin{table*}
\caption{\label{tab:utensor} Six largest components of moir\'e potential $U_{\alpha\bG_0, \beta \bG_i}(\bkbar)$ on high symmetry points when $\theta=1.085^\circ$ and a corrugated structure as an input. The unit is electron Voltage (eV). Unlike classical BM model which sets $u \neq u'$, our method has a detailed description for higher orders components of moir\'e potential.}
\begin{ruledtabular}
\begin{tabular}{lcccccc}
$U_{\mathrm{A}_{1} \mathrm{A}_{2}}(\bar{\Gamma})$ & 0.085809 & 0.079906 & 0.074568 & 0.010323 & 0.010078 & 0.009875 \\
$U_{\mathrm{A}_{1} \mathrm{B}_{2}}(\bar{\Gamma})$ & 0.103212 & 0.097788 & 0.092846 & 0.007722 & 0.007437 & 0.007261 \\
$U_{\mathrm{B}_{1} \mathrm{A}_{2}}(\bar{\Gamma})$ & 0.103212 & 0.097788 & 0.092846 & 0.007722 & 0.007437 & 0.007261 \\
$U_{\mathrm{B}_{1} \mathrm{B}_{2}}(\bar{\Gamma})$ & 0.085809 & 0.079906 & 0.074568 & 0.010323 & 0.010078 & 0.009874 \\
\hline
$U_{\mathrm{A}_{1} \mathrm{A}_{2}}(\bar{\mathrm{K}})$    & 0.080045 & 0.080045 & 0.080045 & 0.010090 & 0.010090 & 0.010090 \\
$U_{\mathrm{A}_{1} \mathrm{B}_{2}}(\bar{\mathrm{K}})$    & 0.097890 & 0.097890 & 0.097890 & 0.007460 & 0.007460 & 0.007460 \\
$U_{\mathrm{B}_{1} \mathrm{A}_{2}}(\bar{\mathrm{K}})$    & 0.097890 & 0.097890 & 0.097890 & 0.007460 & 0.007460 & 0.007460 \\
$U_{\mathrm{B}_{1} \mathrm{B}_{2}}(\bar{\mathrm{K}})$    & 0.080045 & 0.080045 & 0.080045 & 0.010089 & 0.010089 & 0.010089 \\
\hline
$U_{\mathrm{A}_{1} \mathrm{A}_{2}}(\bar{\mathrm{K}'})$    & 0.085737 & 0.080045 & 0.074502 & 0.010300 & 0.010123 & 0.009853 \\
$U_{\mathrm{A}_{1} \mathrm{B}_{2}}(\bar{\mathrm{K}'})$    & 0.103157 & 0.097890 & 0.092799 & 0.007676 & 0.007527 & 0.007217 \\
$U_{\mathrm{B}_{1} \mathrm{A}_{2}}(\bar{\mathrm{K}'})$    & 0.103157 & 0.097890 & 0.092799 & 0.007676 & 0.007527 & 0.007217 \\
$U_{\mathrm{B}_{1} \mathrm{B}_{2}}(\bar{\mathrm{K}'})$    & 0.085737 & 0.080045 & 0.074502 & 0.010300 & 0.010123 & 0.009853 \\
\hline
$U_{\mathrm{A}_{1} \mathrm{A}_{2}}(\bar{\mathrm{M}})$      & 0.082864 & 0.080062 & 0.077246 & 0.010195 & 0.010103 & 0.009972 \\
$U_{\mathrm{A}_{1} \mathrm{B}_{2}}(\bar{\mathrm{M}})$      & 0.100496 & 0.097903 & 0.095316 & 0.007565 & 0.007488 & 0.007336 \\
$U_{\mathrm{B}_{1} \mathrm{A}_{2}}(\bar{\mathrm{M}})$    & 0.100496 & 0.097903 & 0.095316 & 0.007565 & 0.007488 & 0.007336 \\
$U_{\mathrm{B}_{1} \mathrm{B}_{2}}(\bar{\mathrm{M}})$    & 0.082864 & 0.080062 & 0.077246 & 0.010195 & 0.010103 & 0.009971
\end{tabular}
\end{ruledtabular}
\end{table*}

\section{High Order Fourier Components of Moir\'e Potential}\label{app:table}

We tabulate the six largest Fourier components of the moir\'e potential $U(\br)$ for MATBG ($\theta=1.085^\circ$) at high symmetry $\bkbar$ points using our TAPW method with Slater Koster TB parameters in Table.{\ref{tab:utensor}}. TAPW method can help extract effective parameters in generalized BM model by reading these Fourier components.

\begin{figure*}
\includegraphics[width=0.88\textwidth]{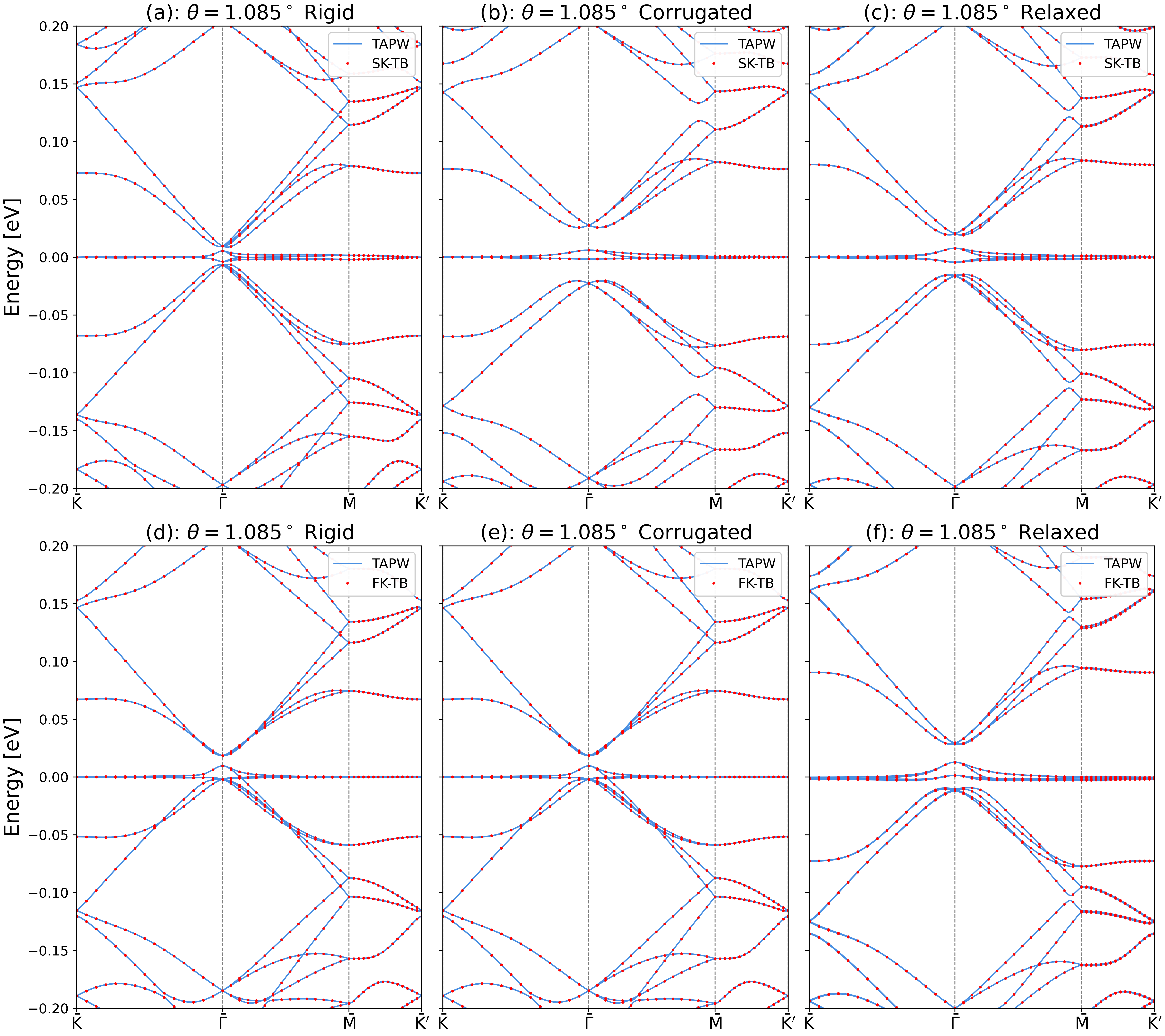}
\caption{\label{fig:all_tb} Electronic Band structures of MATBG ($\theta=1.085^\circ$) along high symmetry points $\bar{\bK}-\bar{\mathbf{\Gamma}}-\bar{\mathbf{M}}-\bar{\bK'}$ using different methods and different TB parameters. (a)-(c): Band structures computed using SK-TB model for rigid, corrugated and relaxed structure, respectively.
(d)-(f):  Band structures computed using FK-TB model for rigid, corrugated and relaxed structure, respectively. The full TB reference results are denoted in red dots while the TAPW results are in blue lines.}
\end{figure*}

\section{More Electronic Band Structure Results for MATBG} \label{app:more_tb}

We plot single particle electronic spectrum of MATBG ($\theta=1.085^\circ$) using different input structures (rigid, corrugated and relaxed TBG structure) with different methods (TAPW and full TB) and TB parameterization (Slater-Koster (SK) TB model and Fang-Kaxiras (FK) TB model \cite{fang2016electronic}) in Fig.~\ref{fig:all_tb}. TAPW results all present good agreement with the full TB results. Lattice relaxation is performed using molecular dynamics with the force field developed in Appx.\ref{app:ff}. The FK-TB parameterization for twisted graphene systems is also integrated in our \texttt{Python} package.

For FK-TB model, which is more accurate according to recent DFT research \cite{pathak2022accurate}, the intra-layer hopping integral $t_{\text{intra}}$ is determined by \cite{kang2022pseudo,vafek2022continuum}
\begin{equation}
    t_{\text{intra}}(\mathbf{r})=t_0 \re^{-\alpha_0\bar{r}^2}\cos(\beta_0\bar{r})+t_1\bar{r}^2 \re^{-\alpha_1(\bar{r}-r_1)^2}.
\end{equation}
Different from Slater-Koster description for inter-layer tunneling, which is isotropic, the inter-layer hopping integral $t_{\text{inter}}(\mathbf{r})$ in FK model is determined by \cite{fang2016electronic, pathak2022accurate, kang2022pseudo, vafek2022continuum}

\begin{equation}
\begin{aligned}
       &t_{\text{inter}}(\mathbf{r})\\
       =&V_0(r)+V_3(r)\left(\frac{1}{3}\sum\limits_{\alpha=1}^3\cos(3\theta_{12}^{(\alpha)})+\frac{1}{3}\sum\limits_{\alpha=1}^3\cos(3\theta_{21}^{(\alpha)})\right)\\
       +&V_6(r)\left(\frac{1}{3}\sum\limits_{\alpha=1}^3\cos(6\theta_{12}^{(\alpha)})+\frac{1}{3}\sum\limits_{\alpha=1}^3\cos(6\theta_{21}^{(\alpha)})\right). 
\end{aligned}
\end{equation}
In above equation, \textbf{r} is the two-dimensional (projected) vector connecting two carbon atoms, $r = |\mathbf{r}|$ and $\bar{r} = r/a$ where graphene lattice constant $a = 2.46 $\r{A}. Note that, FK description for inter-layer hopping cannot capture the effect of lattice corrugation. $V_i(r)$ are fitted as:
\begin{equation}
    \begin{aligned}
        V_0(r)&=\lambda_0 \re^{-\xi_0\bar{r}^2}\cos(\kappa_0\bar{r}),\\
      V_3(r)&=\lambda_3\bar{r}^2 \re^{-\xi_3(\bar{r}-x_3)^2},\\
      V_6(r)&=\lambda_6 \re^{-\xi_6(\bar{r}-x_6)^2}\sin(\kappa_6\bar{r}),
    \end{aligned}
\end{equation}
and $\theta_{12}$ ($\theta_{21}$) indicates the angle between the projected inter-layer bond $\mathbf{r}$ and nearest neighbour bond of atom-1 (atom-2). $\alpha$ is a bond index. All fitted parameters for FK-TB model is summarized in Table.~\ref{tab:fk} \cite{fang2016electronic, pathak2022accurate, kang2022pseudo, vafek2022continuum}.

\begin{table}[b]
\caption{\label{tab:fk}%
Parameters in FK-TB model for TBG.
}
\begin{ruledtabular}
\begin{tabular}{ccccccc}
 Intra &$t_0$ (eV) & $\alpha_0$ & $\beta_0$ & $t_1$ (eV) & $\alpha_1$
 & $r_1$ \\
 \colrule
  & -18.4295 & 1.2771 & 2.3934 & -3.7183 & 6.2194 & 0.9071\\
\hline
 \hline
 Inter & $\lambda_i$ (eV) & $\xi_i$ & $x_i$ & $\kappa_i$  & &\\ 
 \hline
 $V_0$ & 0.3155 & 1.7543 & & 2.0010 & & \\  
 $V_3$ & -0.0688 & -0.0688 & 0.5212  & &\\
 $V_6$ & -0.0083 & 2.8764 & 1.5206 & 1.5731 & &\\
\end{tabular}
\end{ruledtabular}
\end{table}

\section{Details on the Force Field}\label{app:ff}

The potential energy for TBG can be expressed as a summation of the bonded intralayer interactions and non-bonded interlayer interactions.
\begin{equation}
    U= U_{\mathrm{bonded}}+U_{\mathrm{non-bonded}},
\end{equation}
where the bonded interactions can be modelled as $U_{\mathrm{bonded}} = U_2 + U_3 + U_4$ by the Dreiding potential \cite{mayo1990dreiding}, which includes the bond stretch $U_2$ (two-body term), angle bend $U_3$ (three-body term), and dihedral torsion $U_4$ (four-body term). 

Specifically, the bond stretch interactions between carbon $i$ and carbon $j$ can be described by a simple harmonic oscillator as  $U_2^{ij} = \frac{1}{2}k_{ij}(R-R_0)^2 $, where $R_0$ is the equilibrium bond length (i.e. 1.42 $\mathrm{\AA}$  for graphene), and $k_{ij}$ is a constant set as 700 kcal/(mol $\cdot$ \AA$^2$). The three-body angle bend formed by two bonds of atoms $ij$ and $jk$ which share a common atom $j$ can be expressed in a harmonic cosine form as $U_3^{ijk} = \frac{1}{2}C_{ijk}(\cos \theta_{ijk}-\cos \theta_j^0)^2$. In the equation,  $\theta_{ijk}$ is the angle between bonds $ij$ and $jk$, and $\theta_{j}^0 = 120^{\circ}$ is an equilibrium angle. $C_{ijk}$ is a constant set as 133.33 kcal/(mol $\cdot$ rad$^2$). The four-body dihedral interactions $ijkl$ which consists of two bonds $ij$ and $kl$ connected via a common bond $jk$ can be described by  following form as: $U_4^{ijkl} = \frac{1}{2}V_{jk}\{ 1-\cos [n_{jk}(\varphi-\varphi_{jk}^0)]\}$, where $\varphi$ is the dihedral angle between $ijk$ and $jkl$ planes and $\varphi_{jk}^0=180^{\circ}$ is the equilibrium dihedral angle. $n_{jk}$ is the periodicity, which is set as 2, and $V_{jk}$ is the barrier to rotation, which is set as 5 kcal/mol. 

The non-bonded van der Waals interactions $U_{\mathrm{non-bonded}}$  between interlayer TBG can be expressed by the exponential-6 (X6) form \cite{pascal2010quantum} as $U_{\mathrm{vdW}}^{\mathrm{X6}}=A\e^{-R/c}-BR^{-6}$, where $R$ is the distance between two interlayer atoms. The parameters $A$, $c$ and $B$ are set as 385631.5 kcal/mol, $0.2343$ \AA \,and 303.82 kcal/(mol$\cdot $\AA$^6$).  The X6 form has more accurate description regarding the short-range interactions \cite{pascal2010quantum}.

\newpage
\bibliography{apssamp}

\end{document}